\documentclass[]{aaV7.0}

\usepackage{longtable}
\usepackage{graphicx}
\newcommand{\kmprs}  {\mbox{\rm km\,s$^{-1}$}}
\newcommand{\micro}{\mbox{$\xi_{\rm turb}$}}

\newcommand{\feh} {\mbox{\rm [Fe/H]}}
\newcommand{\fehI} {\mbox{\rm [Fe/H]$_{\rm I}$}}
\newcommand{\fehII} {\mbox{\rm [Fe/H]$_{\rm II}$}}

\newcommand{\ofe} {\mbox{\rm [O/Fe]}}
\newcommand{\nafe} {\mbox{\rm [Na/Fe]}}
\newcommand{\mgfe} {\mbox{\rm [Mg/Fe]}}

\newcommand{\cufe} {\mbox{\rm [Cu/Fe]}}

\newcommand{\nife} {\mbox{\rm [Ni/Fe]}}
\newcommand{\znfe} {\mbox{\rm [Zn/Fe]}}

\newcommand{\bay} {\mbox{\rm [Ba/Y]}}
\newcommand{\alphafe} {\mbox{\rm [$\alpha$/Fe]}}
\newcommand{\teff}  {\mbox{$T_{\rm eff}$}}
\newcommand{\logteff} {\mbox{${\rm log}\,T_{\rm eff}$}}
\newcommand{\logg}  {\mbox{{\rm log}\,$g$}}

\newcommand{\FeI} {\ion{Fe}{i}}
\newcommand{\FeII} {\ion{Fe}{ii}}

\newcommand{\chitwored}{$\chi^2_{\rm red}$}
\newcommand{\Mv} {\mbox{$M_V$}}
\newcommand{\Vrot}   {\mbox{$V_{\rm rot}$}}

\def\ltsima{$\; \buildrel < \over \sim \;$}
\def\simlt{\lower.5ex\hbox{\ltsima}}
\def\gtsima{$\; \buildrel > \over \sim \;$}
\def\simgt{\lower.5ex\hbox{\gtsima}}
\def\kms{{\rm\,km\,s$^{-1}$}}
\def\kmskpc{{\rm\,km\,s^{-1}{kpc}^{-1}}}

\def\mathnew{\mathsurround=0pt}
\def\simov#1#2{\lower .5pt\vbox{\baselineskip0pt
    \lineskip-.5pt\ialign{$\mathnew#1\hfil##\hfil$\crcr#2\crcr\sim\crcr}}}

\def\'#1{\ifx#1i{\accent"13\i}\else{\accent"13#1}\fi}

\begin{document}

\title{Two distinct halo populations in the solar neighborhood}

\subtitle{III. Evidence from stellar ages and orbital parameters
\thanks{Based on observations made with the Nordic Optical Telescope
on La Palma, and on data from the European Southern Observatory
ESO/ST-ECF Science Archive Facility (programmes 65.L-0507,
67.D-0086, 67.D-0439, 68.D-0094, 68.B-0475, 69.D-0679, 
70.D-0474, 71.B-0529, 72.B-0585, 76.B-0133 and  77.B-0507).}
\fnmsep\thanks{Tables 1 and 4 are available in electronic form at
{\tt http://www.aanda.org}.}}

\author{W.~J.~Schuster \inst{1}, E.~Moreno \inst{2}, P.~E.~Nissen \inst{3}, \and
B.~Pichardo \inst{2}}

\offprints{W.J.~Schuster}

\institute{
Observatorio Astron\'{o}mico Nacional, Universidad Nacional Aut\'{o}noma
de M\'{e}xico, Apartado Postal 877, C.P. 22800 Ensenada, B.C., M\'{e}xico.
\email{schuster@astrosen.unam.mx}
\and Instituto de Astronom\'{\i}a, Universidad Nacional Aut\'{o}noma de M\'{e}xico,
A.P. 70-264, 04510, M\'{e}xico, D.F., M\'{e}xico.
\email{edmundo@astroscu.unam.mx; barbara@astroscu.unam.mx}
\and Department of Physics and Astronomy, University of Aarhus, DK--8000
Aarhus C, Denmark. 
\email{pen@phys.au.dk}
}

\date{Received  / Accepted }

\abstract
{In Papers I and II of this series, we have found
clear indications of the existence of two distinct populations of stars
in the solar neighborhood belonging to the metal-rich end of the
halo metallicity distribution function.  Based on high-resolution,
high S/N spectra, it is possible to distinguish between 'high-alpha'
and 'low-alpha' components using the \alphafe\ versus \feh\ diagram.}
%
{Precise relative ages and orbital parameters are determined for 67 halo
and 16 thick-disk stars having metallicities in the range $-1.4 <$ \feh $< -0.4$
to better understand the context of the two halo populations in the formation
and evolution of the Galaxy.} 
%
{Ages are derived by comparing the positions of stars in the \logteff --\logg\
diagram with isochrones from the Y$^2$ models interpolated to the exact \feh\
and \alphafe\ values of each star. The stellar parameters have been adopted from
the preceding spectroscopic analyses, but possible systematic errors in \teff\
and \logg\ are considered and corrected.  With space velocities from Paper I as
initial conditions, orbital integrations have been carried out using a detailed,
observationally constrained \object{Milky Way} model including a bar and spiral arms.}
%
{The `high-alpha' halo stars have ages 2--3 Gyr larger than the `low-alpha' ones,
with some probability that the thick-disk stars have ages intermediate between
these two halo components.  The orbital parameters show very distinct differences
between the `high-alpha' and `low-alpha' halo stars. The `low-alpha' ones have
${\rm r}_{\rm max}$'s to 30--40 kpc, ${\rm z}_{\rm max}$'s to $\approx 18$ kpc,
and ${\rm e}_{\rm max}$'s clumped at values greater than 0.85, while the
`high-alpha' ones, ${\rm r}_{\rm max}$'s to about 16 kpc, ${\rm z}_{\rm max}$'s
to 6--8 kpc, and ${\rm e}_{\rm max}$ values more or less uniformly distributed
over 0.4--1.0. }
%
{A dual $in \: situ$-plus-accretion formation scenario best explains the existence and
characteristics of these two metal-rich halo populations, but one remaining defect
is that this model is not consistent regarding the ${\rm r}_{\rm max}$'s obtained for the
$in \: situ$ `high-alpha' component; the predicted values are too small.  It appears that
\object{$\omega$ Cen} may have contributed in a significant way to the existence of the `low-alpha'
component; recent models, including dynamical friction and tidal stripping, have produced
results consistent with the present mass and orbital characteristics of \object{$\omega$ Cen},
while at the same time including extremes in the orbital parameters as great as those of
the `low-alpha' component.}


\titlerunning{Two distinct halo populations.}

\maketitle

\section{Introduction}
\label{introduction}
A stellar population is characterized by the distribution in space, kinematics,
age, and chemical composition of its members.  Probably, the stars in a population
have a common origin and history.  Unraveling the various Galactic populations is 
therefore of high importance for understanding the formation and
evolution of the Galaxy.  In this context, it has been much discussed if the
Galactic halo consists of more than one population.  In the classic confrontation
of early Galactic models, the monolithic collapse model of Eggen, Lynden-Bell
\& Sandage (\cite{eggen62}, ELS) corresponded to a single halo population,
while from a study of globular cluster horizontal branches, Searle and Zinn
(\cite{searle78}, SZ) suggested that the outer globular clusters have a wider range
in ages than the inner ones, and that they represent accretion events which
continued for some time after the inner collapse.  Later, Zinn (\cite{zinn93}) proposed
that the Galactic halo consists of two distinct populations: $i)$ an inner, old,
flattened one with a slight prograde rotation, and $ii)$ an outer, younger,
spherical one with no rotation.  In both these studies stellar ages, or their
proxies, were used to derive the main conclusions.  ELS noted that the most extreme
halo stars, by metallicity, have very elongated, highly elliptical orbits, and so
must have formed during a very rapid Galactic collapse, over an interval of time
short compared to a Galactic rotation period, $\approx 2\times 10^8$ years.  On the
other hand, SZ argued that the second parameter of the globular cluster horizontal
branch morphologies is age, and observed that the outer globular clusters have a
much wider range in color differences along their horizontal branches, implying a
much wider range in ages, $\ga 10^9$ years, than for the inner globular clusters.
(It has been widely discussed whether this second parameter is in fact the cluster
age, but the most recent, more robust conclusions do support this conjecture, for
example, Dotter et al.~2010.)  Many later studies have seen the need to combine
these two scenarios, ELS plus SZ, to more completely describe and understand the
differing kinematics and ages of the inner and outer components of the halo (for
example, Gilmore et al.~\cite{gilmore89}; Zinn \cite{zinn93}; M\'arquez \&
Schuster \cite{marquez94}; Jofr\'e \& Weiss \cite{Jofre11}).  However, the relative
and absolute ages, from color-magnitude and color-metallicity diagrams for 1533
high-velocity and metal-poor stars in the solar neighborhood, have been interpreted
consistently by Schuster et al.~(2006) using the $\Lambda$CDM hierarchical-clustering
scheme for the formation of galaxies without invoking these older scenarios of ELS
and SZ.

The dichotomy of the Galactic halo has been
supported by Carollo et al. (\cite{carollo07}, \cite{carollo10}) from a
study of space motions and metallicities of $\simeq 17\,000$ stars
within 4\,kpc from the Sun in the SDSS survey.  They find that the
inner halo comprises stars with a peak metallicity at $\feh \simeq -1.6$
whereas the outer halo stars distribute around $\feh \simeq -2.2$ with
a net retrograde rotation.  Sch\"{o}nrich et al. (\cite{schonrich11}) have
expressed doubts about these results, claiming the existence of large distance
biases in Carollo et al.~(\cite{carollo10}), and questioning the existence of a
separate outer, metal-poor halo component.  Beers et al.~(\cite{beers11}) have,
however, made a new analysis of the SDSS stars with an improved luminosity
classification, which supports the case for a dual halo.  Also, an independent
mapping of \object{Milky Way} structure based on color-magnitude diagram fitting
of SEGUE photometric data by de Jong et al.~(\cite{dejong10}) provides clear
evidence for a shift in the mean metallicity of the \object{Milky Way}'s stellar
halo, from a peak of [Fe/H] $\sim -1.6$ within 15 kpc to [Fe/H] $\sim -2.2$ at
larger Galactocentric distances, clearly supporting the Carollo et al. dual-halo
view.

In addition, the inner halo itself may, however, consist of more than
one population.  In a study of a local sample of red giant, red
horizontal branch, and RR Lyrae stars with $\feh < -1.0$, Morrison et al. 
(\cite{morrison09}) find evidence of a highly flattened ($c/a \sim 0.2$)
halo component in addition to a moderately flattened
($c/a \sim 0.6$) halo. This highly flattened population is
mainly pressure supported; the mean prograde rotation is 
$\Vrot \simeq 45$\,\kmprs , which distinguishes it from the thick
disk that has $\Vrot \simeq 180$\,\kmprs .  Stars belonging to the
moderately flattened halo have a mean rotation near zero and
a clumpy distribution in energy and angular momentum,
possible remnants of the early accretion of satellite galaxies
(see review by Helmi, \cite{helmi08}).

Another study also makes use of halo stars from the SDSS survey to support
a dual formation scenario for the Galactic halo, that of Jofr\'e \& Weiss
(\cite{Jofre11}).  They derive temperatures and metallicities for a sample
of about 100,000 SDSS stars, and use a Sobel Kernel technique to detect the
turn-offs of halo main sequence stars in the T$_{\rm eff}$ vs \feh\ diagram.
They find ``excellent'' agreement with the $(b$--$y)_{\rm 0,TO}$ versus \feh\
diagram of Schuster et al. (\cite{schuster06}; Figure 9) from $uvby$--$\beta$
photometry, and find clear evidence for a dominating halo population which
formed 10--12 Gyr ago, with no gradient in age and metallicity, and with an
age scatter less than 2 Gyr.  Jofr\'e \& Weiss (\cite{Jofre11}) also find a
large number of halo stars bluer than these turn-offs for $\feh \ga -1.6$,
and suggest that these have come from smaller galaxies accreted later by the
\object{Milky Way}.  In essence they support the combined ELS plus SZ scenario of a
rapid collapse of a proto-galactic cloud to form the inner halo, plus
``collisions and mergers'' to form the outer halo.

In addition to the study of halo field stars, the study of globular clusters
has also shown clear evidence for a halo duality.  A prime example of this
is found in the paper by Mar\'in-Franch et al. (\cite{marinfranch09}) who
obtained precise relative ages for 64 Galactic globular clusters from the deep,
homogeneous photometric data of the ACS Survey of such globular clusters.  They
compared relative positions of the clusters' main-sequence turnoffs, obtained
formal precisions in the relative ages of 2--7\%, and detected two clear
components in the Galactic halo for metallicities $\feh > -1.6$.  The
older group shows an age dispersion of $\approx 5\%$, no age-metallicity relation,
but with a clear galactocentric-distance--metallicity gradient.  The
younger group does show an age-metallicity relation with young clusters being more
metal-rich than older ones.  The age dispersion of the older group ($\la 0.8$
Gyr) corresponds to that expected for the free-fall time of a homogeneous sphere
with the estimated mass and scale-length of the \object{Milky Way}'s dark matter halo,
while the younger group apparently has formed by a different process over a time
interval as long as $\approx 6$ Gyr, which may be a lower limit to the actual
range.  Mar\'in-Franch et al. (\cite{marinfranch09}) point out that the age
dispersion of the older cluster group ``...is not in contradiction with the
formation from the collapse of a single protosystem...'' as in that model proposed
by ELS, while ``...it is very tempting to argue...'' that the clusters of the
younger group were associated with satellite galaxies that have been captured by
the \object{Milky Way} over a significant time interval.  Mar\'in-Franch et al. (\cite
{marinfranch09}) indicate that the galactocentric-distance--metallicity gradient
of the older group is not easily explained by the $\Lambda$CDM cosmological
scenario, while it is also not easy to understand why all of the younger,
accreted clusters should conform to the same age-metallicity relation, if indeed
accreted with different dwarf galaxies, such as Sagittarius, Monoceros, Canis Major,
and others.

Abundance ratios such as $\alpha/$Fe, are also important tracers of
stellar populations and measures of difference in their ages.
Thus, Freeman \& Bland-Hawthorn (\cite{freeman02}) discuss
the possibility to use abundances (chemical tagging) to probe 
the satellite galaxies or proto-clouds from which the Galaxy was assembled
according to the present paradigm of hierarchical structure formation
in CDM cosmologies. The ratio \alphafe , can be
used as a `clock' to probe the star formation rate for the chemical
evolution of a Galactic region. since $\alpha$ refers to
typical alpha-capture elements like Mg, Si, Ca, and Ti, which are mainly
produced during Type II SNe explosions of massive stars on a short time-scale
($\sim 10^7$ years), whereas iron is also produced by Type Ia SNe on a
much longer time scale ($\sim 10^9$ years).

Stellar age is one of the parameters used to identify a stellar population, but
is also that parameter most difficult to measure accurately and absolutely for
individual field stars.  Stellar ages have been used successfully to measure
differences between the inner and outer halo, to compare the thick disk
and halo, and to interpret the Galactic halo in terms of WMAP and
$\Lambda$CDM models (Nissen \& Schuster \cite{nissen91}; M\'arquez \& Schuster
\cite{marquez94}; Schuster et al.~\cite{schuster06}; Jofr\'e \& Weiss \cite{Jofre11}.)
For old field stars it is difficult to obtain the ages much better than 20\%
(Gustafsson \& Mizuno-Wiedner \cite{gustafsson01}), but relative ages within
an ensemble of stars can be more precise, perhaps better than 10\%, and the mean
relative ages between stellar populations provide the best results considering
the large number of stars, and assuming that one can obtain fairly pure samples
and can control systematic effects as a function of \teff\ and metallicity.

The integration of stellar orbits within realistic Galactic mass models has also
provided information useful for characterizing and comparing stellar
components (Allen et al.\ \cite{allen91}; Schuster \& Allen \cite{schuster97});
stellar ages have been combined with orbital integrations to study a possible
difference in age between the inner and outer halo (M\'arquez \& Schuster
\cite{marquez94}).  Non-axisymmetric, observationally constrained, Galactic
potentials, with and without bars and/or spiral arms, have been used to study
orbital characteristics and chaos, and the destruction rates for Galactic globular
clusters (Pichardo et al.\ \cite{PMM04}; Allen et al.\ \cite {allen06}, \cite{allen08}).

Additional evidence for the existence of two distinct halo populations has recently
been obtained by Nissen \& Schuster (\cite{nissen10}, \cite{nissen11}) (Papers I and II)
from a study of chemical abundances of stars in the solar neighborhood.  These halo
stars were selected from Str\"{o}mgren photometry (Schuster et al.~\cite{schuster06})
to have effective temperatures $5200 < \teff < 6300$\,K, metallicities
$-1.6 < \feh < -0.4$, and total space velocities with respect to the local standard
of rest (LSR) $V_{\rm T} \ge 180$\,\kmprs.  This means that they belong to the
metal-rich end of the halo metallicity distribution; halo stars with $\feh < -1.6$
are not represented in the sample.  In addition spectroscopic
data for 16 thick-disk stars were obtained, analyzed, and used for comparison.
The two halo groups  are clearly separated in the \mgfe\ versus \feh\ or \alphafe\
versus \feh\ diagrams, where $\alpha$ represents the average abundance for Mg, Si,
Ca, and Ti.  But also differences between these two halo groups have been
noted for the abundances \nafe, \nife, \cufe, \znfe, and \bay .  These two halo
components, `high-alpha' and `low-alpha', have been interpreted in terms of high and
low star formation rates, respectively, so that the first group obtained chemical
enrichment from Type II supernovae (SNeII) only, while the `low-alpha' component received
chemical enrichment from both SNeII and SNeIa; the latter produce much iron diluting
the overabundance of some products from the SNeII.  Similarities of the `low-alpha'
kinematics to the globular cluster \object{$\omega$ Cen} prompted us to compare their
chemical abundances, but only a partial correspondence, for \nife\ and \cufe, was
obtained; \object{$\omega$ Cen} is lacking in stars showing the SNeIa contamination
characteristic of the `low-alpha' halo stars.  This dual halo result was compared to
the chemical abundances for various dSph and dIrr galaxies in the literature, as well
as to several models used to produce Galactic halos.

In this paper, stellar ages are obtained from the Y$^2$ isochrones (Yi et al.~\cite{yi01},
\cite{yi03}; Kim et al.~\cite{kim02}; and Demarque et al.~\cite{demarque04}), and orbital
parameters from integrations using both the axisymmetric and non-axisymmetric,
observationally-constrained, Galactic mass models of Allen \& Santill\'an (\cite{AS91}),
and Pichardo et al.~(\cite{PMM03}, \cite{PMM04}).  These ages and orbital
parameters are used to better understand the differences between these `high-alpha' and
`low-alpha' halo stars, and to determine more clearly how this dual halo has been formed,
by comparing to three different sets of models, or scenarios, in the literature:  the
classic ELS-plus-SZ scenario, the accretion-plus-accretion (dual-accretion) models of
Font et al.~(\cite{font06a}) and of Robertson et al.~(\cite{Robertson05}), and the
$in \: situ$-plus-accretion models of Zolotov et al. (\cite{zolotov09}, \cite{zolotov10})
and of Purcell et al.~(\cite{purcell10}).

In Sect.~2 the derivation of the stellar parameters (\teff\ and \logg) and of their
errors is described; in Sect.~3 the calculation of the stellar ages for the `high-alpha',
`low-alpha', and thick-disk stars, and the mean ages for these different groups are
presented; in Sect.~4, the integration of the Galactic orbits is described, and several
graphs showing orbital parameters versus other orbital parameters, or versus chemical
abundances, are given; finally Sect.~5 sums it all together with a discussion of the
results and the main conclusions, plus epilogue.

\section{Stellar parameters and abundances}
\label{sect:teff-logg}
As described in Sect. \ref{sect:ages}, stellar ages are determined from
isochrones in the \logg\---\logteff\ plane. The precision
of relative ages is therefore closely related to the errors
in the determination of effective temperature and surface gravity.
In this section, we summarize how these two parameters were 
determined in Papers I and II and check the errors by comparing spectroscopic
and photometric values for stars that are unlikely to be
affected by interstellar reddening.

\subsection{Chemical abundances}
\label{sect:standards}
The chemical abundances of our program stars are based on
equivalent widths (EWs) measured from high-resolution,
high-S/N spectra and analyzed  differentially
with respect to two bright thick-disk disk stars,
\object{HD\,22879} and \object{HD\,76932}.
They have distances of 26 and 21\,pc, respectively,
according to their Hipparcos parallaxes (van Leeuwen \cite{leeuwen07}).
Hence, the two stars are so close that their
colors are not affected by interstellar reddening. Effective temperatures
were determined from $b\!-\!y$ and $V\!-\!K$ using calibrations
derived by Ram\'{\i}rez \& Mel\'endez (\cite{ramirez05}), who
applied the infrared flux method (IRFM) to  determine \teff .
$V$ magnitudes and $b\!-\!y$ indices were taken from
Schuster et al.~(\cite{schuster06}) and $K$ magnitudes from the 2MASS catalogue
(Skrutskie et al. \cite{skrutskie06}). The resulting values of \teff\
are given in Table \ref{table:param} (online).
As seen, there is excellent agreement between the temperatures
from $b\!-\!y$ and $V\!-\!K$; the difference is less than 15\,K for both stars.

The surface gravities of \object{HD\,22879} and \object{HD\,76932} given in
Table \ref{table:param} were determined from the fundamental relation
\begin{eqnarray} \log \frac{g}{g_{\odot}}  =  \log \frac{\cal{M}}{\cal{M}_{\odot}} +
4 \log \frac{\teff}{T_{\rm eff,\odot}} + 0.4 (M_{\rm bol} - M_{{\rm bol},\odot})
\end{eqnarray} where $\cal{M}$ is the mass of the star and $M_{\rm bol}$ the absolute
bolometric magnitude. The Hipparcos parallax was used to derive \Mv\
and the bolometric correction adopted from Alonso et al. (\cite{alonso95}).
The stellar mass was obtained by interpolating in the \Mv --\logteff\
diagram between the evolutionary tracks of VandenBerg et al. (\cite{vandenberg00}).
Due to the small error of the Hipparcos parallaxes of \object{HD\,22879} and
\object{HD\,76932}, $\sigma (\pi) / \pi < 0.015$ (van Leeuwen \cite{leeuwen07}),
the estimated error of \logg\ is only 0.03\,dex for both stars.

With the so derived values of \teff\ and \logg , chemical abundances
relative to the Sun were derived for the two standard stars using a
subset of spectral lines for which the equivalent widths
could be measured reliably in the solar flux spectrum
(Kurucz et al. \cite{kurucz84}). Adopting these
abundances, an `inverted' analysis led to the determination of
$gf$-values for all lines. The derived $gf$-values for
\object{HD\,22879} and \object{HD\,76932} agree within $\pm
0.03$\,dex for the large majority of lines; hence, for each line,
the mean value of $gf$ (see Table 3 in Paper II) were adopted and used for an LTE
analysis of all program stars based on MARCS model atmospheres from
Gustafsson et al. (\cite{gustafsson08}). This method
ensures that high-precision differential abundances relative to
the two standard stars are obtained. Further details about spectral
lines and line-broadening mechanisms included are given in
Papers I and II.

\subsection{Effective temperature}
\label{sect:teff}
For the majority of the 94 program stars, interstellar NaD lines
are clearly seen in their spectra (see Tables 1 and 2 in Paper I). The
colors of these stars are therefore probably affected by interstellar reddening.
Although, the reddening excess may be estimated via the H$\beta$
index (Schuster \& Nissen \cite{schuster89}) or from the
strength of the NaD lines (Alves-Brito et al. \cite{alves-brito10}),
the precision of \teff\ derived from colors will not be as high as in
the case of unreddened stars.  Therefore, we preferred to determine
effective temperatures spectroscopically, i.e.  by requiring that
the Fe abundances derived from \FeI\ lines do not depend systematically on
excitation potential.

\begin{figure}
\resizebox{\hsize}{!}{\includegraphics{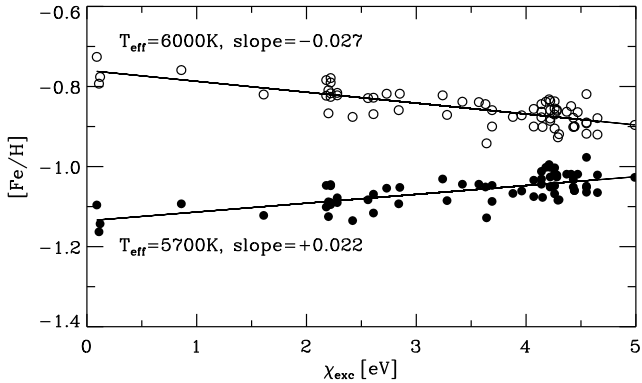}}
\caption{[Fe/H] values derived from \FeI\ lines with EW\,$< 50$\,m\AA\ in
the spectrum of \object{G\,56-30} as a function of excitation potential
for the lower energy level. The open circles refer to an
effective temperature of 6000\,K for the model atmosphere used and the filled
circles to a model with \teff\ = 5700\,K.}
\label{fig:exc.pot}
\end{figure}

Fig. \ref{fig:exc.pot} illustrates the method in the case of
\object{G\,56-30}.  By interpolating between the slopes for \teff\ = 5700\,K
and 6000\,K, an effective temperature of 5830\,K is derived for this star. As the
\FeI\ lines are also applied to determine the microturbulence, \micro , by
minimizing the dependence of \feh\ on equivalent width, only
lines with EW\,$< 50$\,m\AA\ were used to estimate \teff , whereas
\micro\ is based only on \FeI\ lines with $\chi_{\rm exc} > 3.0$\,eV. In this
way, the determination of \teff\  and \micro\ is to some extent decoupled,
although it is necessary to iterate in order to obtain
consistent values of \teff\  and \micro .  From the error
of the excitation slopes, we estimate that the 1-$\sigma$ statistical error
of \teff\ is on the order of 20 to 30\,K depending on the number
of \feh\ lines available.


A check of the estimated precision of the effective
temperatures as derived from the excitation balance of \FeI\ lines
may be obtained by comparing with temperatures derived from
$b\!-\!y$ and $V\!-\!K$ for  stars having no detectable
interstellar NaD lines, i.e.\ EW\,$\simlt 3$\,m\AA\ for UVES spectra and
EW\,$\simlt 5$\,m\AA\ for FIES spectra. Such stars
are unlikely to be affected by interstellar reddening
(e.g. Nissen \cite{nissen94}; Alves-Brito et al.  \cite{alves-brito10}).
\teff\ values as derived from $b\!-\!y$ and $V\!-\!K$ using
the Ram\'{\i}rez \& Melend\'{e}z (\cite{ramirez05}) calibrations and
the same sources of photometry as in the case of the standard stars
are given in Table \ref{table:param} together with the spectroscopic values
of \teff .  As listed in the last column of the table, four of
the stars are single-lined spectroscopic binaries and another four stars have
close companions, which are probably affecting the color indices and hence the
derived values of \teff .


\onltab{1}{
\begin{table*}
\caption[ ]{Spectroscopic and photometric values of \teff\ and \logg\ for
stars without detectable interstellar NaD lines.}
\label{table:param}
\setlength{\tabcolsep}{0.2cm}
\begin{tabular}{lcccccccl}
\noalign{\smallskip}
\hline\hline
\noalign{\smallskip}
\noalign{\smallskip}
  ID & \teff (spec)  & \logg (spec)  & \feh  & \alphafe  & \teff ($b\!-\!y$)  & \teff ($V\!-\!K$)  & \logg (HIP)  &   Comment \\
     &   [K]         &  [cgs]        &       &           &    [K]         &      [K]       &   [cgs]      &   \\
\noalign{\smallskip}
\hline
\noalign{\smallskip}
BD$-$21 3420 &  5808 &  4.26 & $-$1.13 &  0.31 &  5855 &  5884 &  4.09\,$\pm 0.18$ &  \\
CD$-$33 3337 &  5979 &  3.86 & $-$1.36 &  0.30 &  6005 &  5924 &  4.08\,$\pm 0.08$ &  \\
CD$-$45 3283 &  5597 &  4.55 & $-$0.91 &  0.12 &  5646 &  5616 &  4.71\,$\pm 0.13$ &  \\
CD$-$57 1633 &  5873 &  4.28 & $-$0.90 &  0.07 &  5964 &  5874 &  4.26\,$\pm 0.08$ &  \\
G13-38     &  5263 &  4.54 & $-$0.88 &  0.32 &  5241 &  5279 &  4.45\,$\pm 0.12$ &  \\
G18-28     &  5372 &  4.41 & $-$0.83 &  0.31 &  5376 &  5320 &  4.58\,$\pm 0.08$ & SB1 (1)  \\
G46-31     &  5901 &  4.23 & $-$0.83 &  0.15 &  5918 &  5811 &  4.15\,$\pm 0.40$ & SB1 (1)  \\
G56-30     &  5830 &  4.26 & $-$0.89 &  0.11 &  5871 &  5824 &                   &  \\
G82-05     &  5277 &  4.45 & $-$0.75 &  0.09 &  5312 &  5293 &  4.73\,$\pm 0.14$ &  \\
G85-13     &  5628 &  4.38 & $-$0.59 &  0.28 &  5631 &  5681 &  4.31\,$\pm 0.20$ &  \\
G94-49     &  5373 &  4.50 & $-$0.80 &  0.31 &  5346 &  5363 &                   &  \\
G99-21     &  5487 &  4.39 & $-$0.67 &  0.29 &  5528 &  5536 &  4.31\,$\pm 0.27$ &  \\
G119-64    &  6181 &  4.18 & $-$1.48 &  0.28 &  6129 &  6165 &  4.07\,$\pm 0.15$ &  \\
G121-12    &  5928 &  4.23 & $-$0.93 &  0.10 &  5913 &  5919 &  4.17\,$\pm 0.24$ &  \\
G159-50    &  5624 &  4.37 & $-$0.93 &  0.31 &  5622 &  5660 &  4.29\,$\pm 0.08$ &  \\
G172-61    &  5225 &  4.47 & $-$1.00 &  0.19 &  5282 &  5242 &                   & SB1 (1) \\
G176-53    &  5523 &  4.48 & $-$1.34 &  0.18 &  5522 &  5595 &  4.36\,$\pm 0.10$ &  \\
G180-24    &  6004 &  4.21 & $-$1.39 &  0.33 &  5958 &  5986 &  4.18\,$\pm 0.11$ &  \\
G232-18    &  5559 &  4.48 & $-$0.93 &  0.32 &  5505 &  5546 &                   &  \\
HD3567     &  6051 &  4.02 & $-$1.16 &  0.21 &  6027 &  6014 &  4.12\,$\pm 0.10$ &  \\
HD17820    &  5773 &  4.22 & $-$0.67 &  0.29 &  5736 &  5787 &  4.20\,$\pm 0.06$ &  \\
HD22879    &  5759 &  4.25 & $-$0.85 &  0.31 &  5763 &  5754 &  4.25\,$\pm 0.03$ &  \\
HD25704    &  5868 &  4.26 & $-$0.85 &  0.24 &  5752 &  5659 &  4.20\,$\pm 0.04$ &  $\Delta \! H_p=3.49, \rho = 1\farcs10$ (2)\\
HD51754    &  5767 &  4.29 & $-$0.58 &  0.26 &  5769 &  5814 &  4.24\,$\pm 0.08$ &  \\
HD59392    &  6012 &  3.91 & $-$1.60 &  0.32 &  5992 &  5976 &  3.99\,$\pm 0.15$ &  \\
HD76932    &  5877 &  4.13 & $-$0.87 &  0.29 &  5883 &  5871 &  4.13\,$\pm 0.03$ &  \\
HD97320    &  6008 &  4.19 & $-$1.17 &  0.28 &  5979 &  5992 &  4.22\,$\pm 0.04$ &  \\
HD106516   &  6196 &  4.42 & $-$0.68 &  0.29 &  6202 &  6169 &  4.36\,$\pm 0.03$ & SB1 (3) \\
HD114762A  &  5856 &  4.21 & $-$0.70 &  0.24 &  5815 &  5748 &  4.19\,$\pm 0.04$ &  \\
HD120559   &  5412 &  4.50 & $-$0.89 &  0.30 &  5361 &  5352 &  4.56\,$\pm 0.04$ &  \\
HD121004   &  5669 &  4.37 & $-$0.70 &  0.32 &  5596 &  5601 &  4.36\,$\pm 0.07$ &  \\
HD126681   &  5507 &  4.45 & $-$1.17 &  0.35 &  5506 &  5497 &  4.61\,$\pm 0.06$ &  \\
HD148816   &  5823 &  4.13 & $-$0.73 &  0.27 &  5809 &  5803 &  4.10\,$\pm 0.04$ &  \\
HD159482   &  5737 &  4.31 & $-$0.73 &  0.30 &  5687 &  5614 &  4.36\,$\pm 0.06$ & $\Delta \! H_p=1.47, \rho = 0\farcs27$ (2) \\
HD175179   &  5713 &  4.33 & $-$0.65 &  0.29 &  5687 &  5719 &  4.31\,$\pm 0.08$ &  \\
HD177095   &  5349 &  4.39 & $-$0.74 &  0.31 &  5321 &  5375 &  4.45\,$\pm 0.08$ &  \\
HD189558   &  5617 &  3.80 & $-$1.12 &  0.33 &  5612 &  5652 &  3.82\,$\pm 0.05$ &  \\
HD193901   &  5656 &  4.36 & $-$1.09 &  0.16 &  5636 &  5743 &  4.47\,$\pm 0.05$ &  \\
HD194598   &  5942 &  4.33 & $-$1.09 &  0.18 &  5936 &  6000 &  4.22\,$\pm 0.05$ &  \\
HD199289   &  5810 &  4.28 & $-$1.04 &  0.30 &  5754 &  5849 &  4.23\,$\pm 0.05$ &  \\
HD205650   &  5698 &  4.32 & $-$1.17 &  0.30 &  5691 &  5808 &  4.44\,$\pm 0.06$ &  \\
HD219617   &  5862 &  4.28 & $-$1.45 &  0.28 &  5887 &  5967 &  4.26\,$\pm 0.11$ & $\Delta \! H_p=0.22, \rho = 0\farcs82$ (2) \\
HD222766   &  5334 &  4.27 & $-$0.67 &  0.30 &  5380 &  5380 &  4.14\,$\pm 0.19$ &  \\
HD230409   &  5318 &  4.54 & $-$0.85 &  0.27 &  5258 &  5336 &  4.49\,$\pm 0.11$ &  \\
HD233511   &  6006 &  4.23 & $-$1.55 &  0.34 &  5932 &  5985 &  4.28\,$\pm 0.14$ &  \\
HD237822   &  5603 &  4.33 & $-$0.45 &  0.29 &  5621 &  5611 &  4.49\,$\pm 0.13$ &  \\
HD241253   &  5831 &  4.31 & $-$1.10 &  0.29 &  5834 &  5897 &  4.15\,$\pm 0.18$ &  \\
HD250792A  &  5489 &  4.47 & $-$1.01 &  0.24 &  5511 &  5363 &  4.37\,$\pm 0.11$ & $\Delta \! H_p=2.11, \rho = 0\farcs22$ (2) \\
\noalign{\smallskip}
\hline
\end{tabular}

\medskip

References:
(1) Latham et al. (\cite{latham02}).
(2) ESA (\cite{esa97}).
(3) Ducati et al. (\cite{ducati11}).

\end{table*}
}

\begin{figure}
\resizebox{\hsize}{!}{\includegraphics{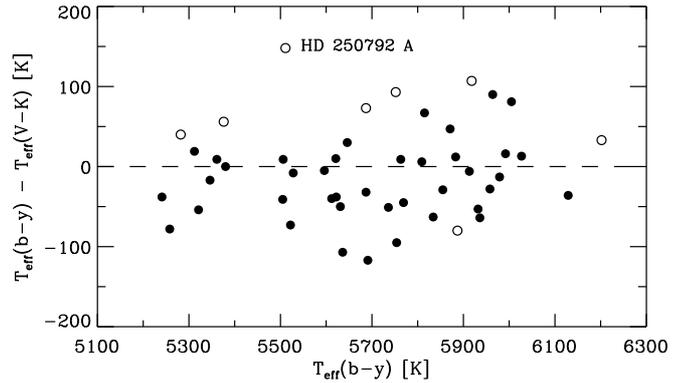}}
\caption{The difference of effective temperatures as determined from
from the photometric colors
$b\!-\!y$ and $V\!-\!K$ for stars with no detectable interstellar
NaD lines. Known binary stars are shown with open circles.}
\label{fig:Teff.b-y.V-K}
\end{figure}

Fig. \ref{fig:Teff.b-y.V-K} shows a comparison of the effective
temperatures derived from $b\!-\!y$ and $V\!-\!K$.
As seen, $\teff (b\!-\!y)$ is larger than $\teff (V\!-\!K)$ for
seven of the eight binary stars. This can be explained if
the secondary component is a red, low-mass star affecting
in particular the infrared $K$ magnitude. An exception
is \object{HD\,219617} for which the two components have
similar magnitudes and nearly identical spectra.
(Takeda \& Takada-Hidai \cite{takeda11}).


There are no signatures of a secondary component in the spectra
of the binary stars, except for the spectrum of \object{HD\,250792\,A},
which has unusually strong and slightly shifted MgH lines around
5100\,\AA\ probably arising from a cool component.
\object{HD\,250792\,A} is the binary star that shows the
largest difference between $\teff (b\!-\!y)$ and $\teff (V\!-\!K)$,
i.e. 148\,K.

\begin{figure}
\resizebox{\hsize}{!}{\includegraphics{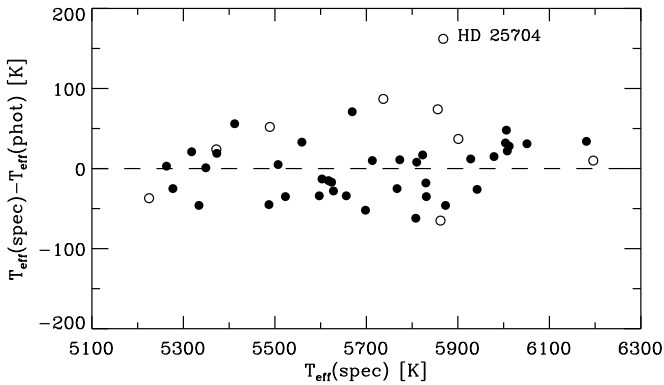}}
\caption{The difference of effective temperatures determined from
the excitation balance of \FeI\ lines and from the photometric colors
$b\!-\!y$ and $V\!-\!K$ for stars without detectable interstellar
NaD lines. Known binary stars are shown with open circles.}
\label{fig:Teff.spec-phot}
\end{figure}

Excluding the binary stars, the rms scatter of the difference
$\teff (b\!-\!y) - \teff (V\!-\!K)$ is 50\,K.
This suggests that each of the two temperatures are determined
with a precision of $\sim$\,35\,K, and that the precision of the
mean photometric temperature, 
\teff (phot) = 1/2 ($\teff (b\!-\!y) + \teff (V\!-\!K)$), is
about 25\,K for single stars. A comparison of this photometric temperature
with the spectroscopic value of \teff\ is shown in
Fig. \ref{fig:Teff.spec-phot} (excluding the two standard stars
for which the two sets of temperatures agree by definition).
As seen, the agreement is excellent.  Without the binary stars,
the rms scatter of the difference is 33\,K. Adopting an
error of 25\,K in \teff (phot) this corresponds to an error
of 22\,K of \teff (spec).  If the binary stars
are included, the rms scatter of the difference between
\teff (spec) and  \teff (phot) increases to 44\,K with a large
contribution to the scatter coming from \object{HD\,25704}.

We suggest that binarity is affecting \teff (phot) more than
it is affecting \teff (spec).
Out of the 94 program stars, eight stars are
designated as SB1 in the SIMBAD database and five stars have close
companions according to the Hipparcos catalogue, but only in
two cases (\object{HD\,250792\,A} and \object{HD\,163810}),
there are signs of a secondary spectrum in our high-resolution spectra,
i.e.  unusually strong and asymmetric MgH lines. For these
two stars, \teff\ for the primary component is
probably slightly higher than the derived \teff (spec).


On the basis of these comparisons, we adopt $\pm 30$\,K 
as the 1-sigma statistical error for the effective temperatures
used in the age determinations.
The systematic error may, however be larger. Our \teff\ values
refer to the IRFM temperatures of Ram\'{\i}rez \& Mel\'endez
(\cite{ramirez05}), but recently Casagrande et al. (\cite{casagrande10})
have carefully revisited the IRFM method for dwarf and subgiant stars
finding a systematic offset of +100\,K in \teff\ relative to the
Ram\'{\i}rez \& Mel\'endez values for stars with $\feh > -2.0$ and
$4800 < \teff < 6200$\,K (see Fig. 5 in Casagrande et al. 
\cite{casagrande10}). The difference mainly stems from the infrared
absolute flux calibrations applied; Casagrande et al. have taken
advantage of accurate HST spectrophotometry of Vega by Bohlin
(\cite{bohlin07}).  Furthermore, they have validated the new \teff\
scale by interferometric angular diameter measurements and solar twins,
i.e. stars having high-resolution spectra indistinguishable from the
solar flux spectrum and therefore \teff\ close to 5780\,K.
Thus, it seems that the \teff\ values published in Papers I and II
are systematically too low by $\sim 100$\,K; as mentioned in
Sect. \ref{sect:ages} we have corrected for this in the age
determinations.

It should be noted that an increase of 100\,K in \teff\
has only a small effect on the abundances derived. This is due
to the fact that \feh\ is determined from \FeII\ lines and that the
abundance ratios are based on lines corresponding to the same ionization
stage of the elements. The correction of \feh\ is about $-0.03$\,dex at
\teff \,= 5400\,K and +0.01\,dex at \teff \,= 6100\,K, whereas
the correction of \alphafe\ is approximately constant at $-0.01$\,dex.
These corrections have a very small effect on the derived ages
and may be neglected.

\subsection{Surface gravity}
\label{sect:gravity}
Several of the program stars are missing Hipparcos parallaxes and
in other cases, the relative error of the parallaxes is too large
to allow a precise determination of \logg\ from Eq. (1). Therefore,
we have preferred to determine a spectroscopic value of \logg\
from the requirement
that the Fe abundance derived from \FeI\ and \FeII\ lines should
have the same difference as in the case of the standard stars, i.e.
\fehII\ $-$ \fehI\ = 0.075.  This difference is probably due to a
departure from LTE in the ionization balance of Fe.  It is therefore
implicitly assumed that this non-LTE deviation does not change
significantly as a function of \teff\ or \feh\ for the
sample of stars considered.

The strength of \FeI\ lines is nearly independent of
\logg , whereas the Fe abundance derived from \FeII\ lines increases
by about 0.04\,dex if \logg\ of the model atmosphere  is
increased by 0.1\,dex. Since we have about 90 \FeI\ and 15 \FeII\
lines available, the difference  \fehII\ $-$ \fehI\ can be determined
to a precision of about 0.02\,dex. This corresponds then to a
precision of 0.05\,dex in the determination of \logg . Again, it
should be emphasized that this estimate refers to differential
values of \logg\ relative to the gravities of the two standard stars.

\begin{figure}
\resizebox{\hsize}{!}{\includegraphics{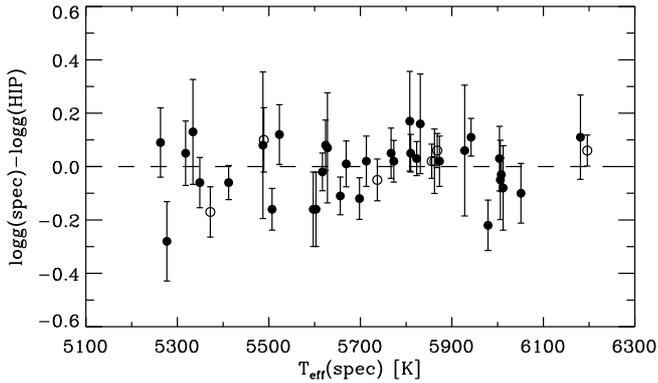}}
\caption{The difference of logarithmic surface gravity
derived from the ionization balance of Fe lines and from
Eq. (1) using Hipparcos parallaxes to determine the absolute bolometric
magnitude. Only stars with no detectable interstellar NaD lines are
included, and known binaries are shown with open circles.}
\label{fig:dellogg}
\end{figure}

As a test of the gravities derived from the ionization balance of
Fe lines, Fig. \ref{fig:dellogg} compares \logg (spec) with the
gravity derived from Eq. (1) using Hipparcos parallaxes to determine
the distance and hence the absolute magnitude.
To avoid effects from reddening only stars without detectable
interstellar NaD lines are included in the comparison, i.e. the
sample listed in Table \ref{table:param}.  The derived values
of \logg (HIP) and their errors (with the main contribution coming
from the error of the parallax) are given in Col. 8. As seen,
four stars have no Hipparcos parallax and one star (\object{G\,46-31})
has a very large error in \logg (HIP). The rest
(except the two standard stars, for which the two sets of gravities
agree by definition) are plotted in Fig. \ref{fig:dellogg}.
The error bars on  $\Delta \logg$ = \logg (spec) $-$ \logg (HIP)
are calculated as a quadratic addition of the errors of \logg (HIP)
and \logg (spec), the latter assumed to be $\pm 0.05$\,dex.

Fig. \ref{fig:dellogg} shows a very satisfactory
agreement between \logg (spec) and \logg (HIP).
The reduced chi-square
\chitwored\,$= 1/N \sum (\Delta \logg_i / \sigma_i)^2$
is 0.94, and there is no significant offset of the binary stars
\footnote{In deriving \logg (HIP) for \object{HD\,219617} we have
used the visual magnitude, $V = 8.77$ of the primary component
as measured by Takeda \& Takada-Hidai (\cite{takeda11}).
For the other three close binaries we have estimated the magnitude
of the primary component by using the magnitude difference given in
Table \ref{table:param}.}.
This suggests that $\pm 0.05$\,dex is a realistic estimate
for the error of \logg (spec).


As discussed in Sect. \ref{sect:teff}, the new IRFM temperatures by
Casagrande et al. (\cite{casagrande10}) suggest that the \teff\ values
in Papers I and II are systematically too low by 100\,K.
This change of the temperature
scale has a small systematic effect on the estimated gravities; due to
the \teff\ term in Eq. (1), the \logg\ values of the standard stars
increase by 0.03\,dex.

\section{Ages}
\label{sect:ages}
The relative ages of this study have been determined using the Y$^2$
isochrones of Yi et al.~(\cite{yi01}, \cite{yi03}), Kim et al.~(\cite{kim02}),
and Demarque et al.~(\cite{demarque04}), which allow interpolation to the
exact values of \feh\ and \alphafe\ as determined by our spectroscopic
measures. \feh\ represents the mean metallicity of the star, and \alphafe\
the deviation of the alpha elements from this mean.  For each of our halo
and thick-disk stars, an isochrone set in the \logg\---\logteff\ plane has
been interpolated to the exact spectroscopic values of \feh\ and \alphafe\
for that star.

However, before the final relative ages could be derived, problems
relating to possible systematic effects had to be resolved. As discussed
in Sect. \ref{sect:teff}, the recent work of Casagrande et al.\
(\cite{casagrande10}) suggests that a correction of +100\,K should be
applied to the spectroscopic temperatures in Papers I and II before 
comparing each star to its corresponding isochrone set.  This systematic
correction lowers the resulting ages to values more in line with the WMAP
age for the Universe ($13.75 \pm0.13$ Gyr; Jarosik et al.~\cite{jarosik11},
Larson et al.~\cite{larson11}).  Also, Jofr\'e \& Weiss \cite{Jofre11}, in
their study of the turn-offs of halo main sequence stars from the SDSS survey,
show that gravitational settling of heavy elements (diffusion) is necessary
for the isochrone models to produce ages compatible with the age of the
Universe; models without this diffusion can produce ages as much as 4 Gyr
larger.  The Y$^2$ isochrones used in the present paper do make use of
such gravitational settling. 

In addition, our 13 coolest stars, $\teff < 5600$ K, show systematic vertical
offsets (in \logg ) with respect to the isochrones in the \logg\---\logteff\
plane.  Such cool stars should be little evolved from their corresponding
ZAMS isochrones, even for the total age of the Universe, and so these 13 stars
(twelve halo and one thick-disk) have been fit to isochrones with ages of 13 Gyr
at their corresponding values of \feh\ and \alphafe\, and with the above
correction of +100 K to the stars' \teff\ values.  Such fitting derives a
correction of $\Delta \logg = -0.127$ dex, which has been applied to all Y$^2$
isochrones used for our age analyses

A small part of this gravity correction stems from the need for a systematic
change  of $+0.03$ dex in the stellar \logg\ values caused by the $+100$ K
increase in \teff\ (Eq.~1).  The remaining part seems to be due to a
systematic error in \logg\ of the isochrones.  A corresponding offset between
isochrones and unevolved stars with metallicities in the range
$-1.0< $[Fe/H]$ <-0.5$ is present in the M$_{\rm bol}$--log \teff\ diagram
(Lebreton \cite{lebreton00}) even if the new \teff\ scale of Casagrande et
al.~(\cite{casagrande10}) is adopted (Casagrande et al. \cite{casagrande11},
Fig.~12).

The \logg\ offset of the isochrones relative to the unevolved stars
may be related to the mixing length parameter, l/H$_{\rm p}$,
used in modeling the upper convection zone. The stellar radius
and hence \logg\ depends on its value.  Usually the mixing length
parameter is assumed to be independent of \feh\ and is determined
by fitting a model of the Sun to the solar parameters, e.g.
l/H$_{\rm p}$ = 1.74 in the case of the Y$^2$ isochrones.  If instead
l/H$_{\rm p}$ decreases to about 1.2 at \feh $ = -1$, the \logg\ offset
of the isochrones would go away according to the calculations of
VandenBerg (\cite{vandenberg83}).  Such a change of l/H$_{\rm p}$ would
also decrease \teff\ of the isochrones in the turnoff region and hence
decrease the derived ages of our stars by 2--3 Gyr.  For this reason,
the absolute ages of our stars are rather uncertain, but the
relative ages of the stars at a given metallicity are
insensitive to the mixing length parameter and should be quite precise.

\begin{figure}
\resizebox{\hsize}{!}{\includegraphics{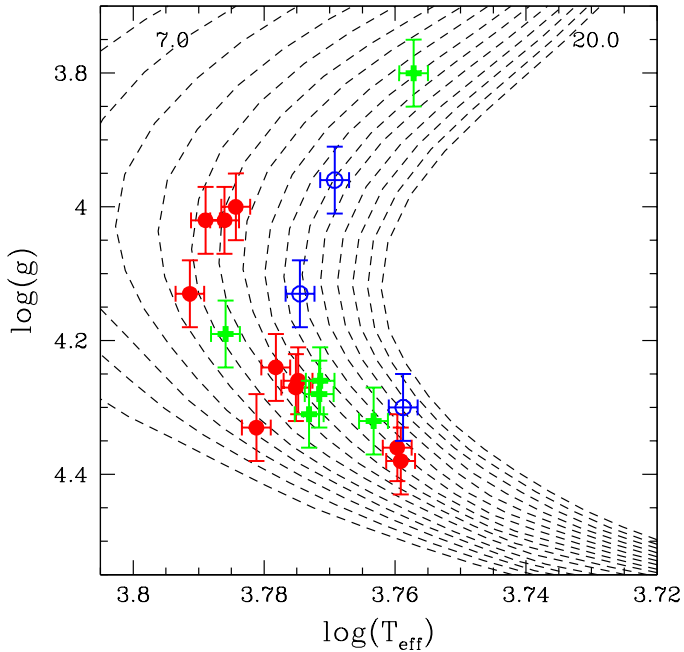}}
\caption{The \logg\---\logteff\  diagram for stars with
$-1.20 < \feh < -0.975$. Y$^2$ isochrones corresponding to
\feh\ = $-1.0875$ and \alphafe\ = +0.25 
are over-plotted in steps of 1 Gyr from 5 to 20\,Gyr.
The full (red) circles show
the  `low-alpha' halo stars, the open (blue) circles the `high-alpha' halo,
and the (green) pluses the thick-disk.}  
\label{fig:logg-logteff1}
\end{figure}

\begin{figure}
\resizebox{\hsize}{!}{\includegraphics{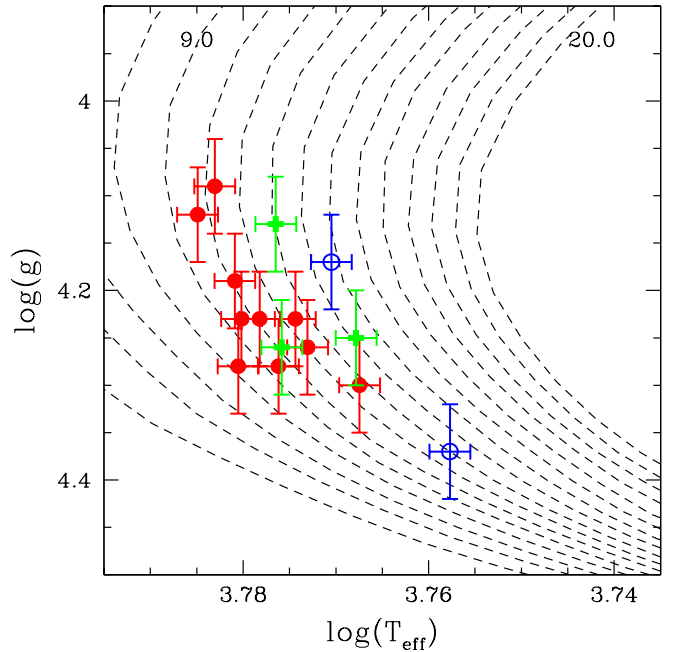}}
\caption{The \logg\---\logteff\  diagram for stars with
$-0.975 < \feh < -0.775$. Y$^2$ isochrones corresponding to
\feh\ = $-0.875$ and \alphafe\ = +0.215 are
over-plotted in steps of 1 Gyr from 5 to 20\,Gyr. The symbols
as in Fig.~\ref{fig:logg-logteff1}.}
\label{fig:logg-logteff2}
\end{figure}

In Figs.~\ref{fig:logg-logteff1} and \ref{fig:logg-logteff2} are shown groups
of `high-alpha' halo, `low-alpha' halo, and thick-disk stars over the metallicity
ranges of $-1.20 < \feh <-0.975$ and $-0.975 < \feh < -0.775$, respectively.
The Y$^2$ isochrones have been interpolated to the corresponding mean values of
\feh\ = $-1.0875$ and \alphafe\ = $+0.250$, and  \feh\ = $-0.875$ and
\alphafe\ = $+0.215$, respectively.  The full (red) circles show the  `low-alpha'
halo stars, the open (blue) circles the `high-alpha' halo, and the (green) pluses
the thick-disk.  The sizes of the error bars are those derived above in 
Subsection~3.6:  $\pm 0.05$ dex in \logg\ and $\pm 0.0022$ in \logteff\
(corresponding to $\pm 30$\,K in \teff ). 
The correction +100\,K has been added to the spectroscopic \teff\
values of the stars, and the isochrones shifted by $\Delta \logg = -0.127$ dex.
Both figures show clear evidence for $\langle$Age$\rangle_{\rm low-alpha} <
\langle$Age$\rangle_{\rm high-alpha}$, and both suggest a sequence:
$\langle$Age$\rangle_{\rm low-alpha} < \langle$Age$\rangle_{\rm thick-disk} <
\langle$Age$\rangle_{\rm high-alpha}$.

In deriving ages for individual stars, we have also set the limits $3.8 < \logg
< 4.4$ for those stars which will have their ages estimated.  Outside these limits
the isochrone spacing is too small to provide a good precision for these age 
determinations.

In Table \ref{table:ave_ages}, mean, weighted ages are given for the four
components:  `high-alpha' halo, `low-alpha' halo, thick-disk, and `high-alpha'
halo plus thick-disk, and for five metallicity ranges from $-1.40 < \feh < -1.20$
to $-0.575 < \feh < -0.40$.  For these weighted averages each star has been
plotted in a separate \logg\---\logteff\ diagram with the Y$^2$ isochrones
interpolated exactly to that star's \feh\ and \alphafe\ values.
The age of each star has been interpolated in its corresponding
\logg\---\logteff\ diagram, as well as the ages at each of the corners of the
error box constructed from the error bars.  Each of these ``corner'' ages
provides an age error estimate for that star, and the four of these have been
averaged for the final error estimate of that star's age, $\sigma_{\star}$.  Then the
weight for that star's age has been taken as:  $1/\sigma_{\star}^2$.  In Table
\ref{table:ave_ages} these weighted ages are given for each of the Galactic
components, over the five metallicity ranges, as well as the weighted standard
deviations, and the number of stars in each group.

The results of this Table \ref{table:ave_ages} are concordant with those from
Figs.~\ref{fig:logg-logteff1} and \ref{fig:logg-logteff2}, showing clearly that
$\langle$Age$\rangle_{\rm low-alpha} < \langle$Age$\rangle_{\rm high-alpha}$,
and suggesting again a sequence:  $\langle$Age$\rangle_{\rm low-alpha} <
\langle$Age$\rangle_{\rm thick-disk} < \langle$Age$\rangle_{\rm high-alpha}$.


\begin{table*}

\caption[ ]{Average, weighted ages (in Gyr) with mean errors and the number
of stars, for five ranges in \feh\ and for the four components
mentioned in the text:  `high-alpha' halo, `low-alpha' halo, thick-disk,
and `high-alpha' halo plus thick-disk}.
\label{table:ave_ages}
\setlength{\tabcolsep}{0.10cm}
\begin{tabular}{lrcrcrcrcrcrcrcrcrcrcr}
\hline\hline
\noalign{\smallskip}
\noalign{\smallskip}
Group / \feh : & \multicolumn{4}{c}{[$-1.40, -1.20$]}
               & \multicolumn{4}{c}{[$-1.20, -0.975$]} 
               & \multicolumn{4}{c}{[$-0.975, -0.775$]}
               & \multicolumn{4}{c}{[$-0.775, -0.575$]}
               & \multicolumn{4}{c}{[$-0.575, -0.40$]} \\
\noalign{\smallskip}
               & $\langle$Age$\rangle$ & $\sigma$ & N &
               & $\langle$Age$\rangle$ & $\sigma$ & N &
               & $\langle$Age$\rangle$ & $\sigma$ & N &
               & $\langle$Age$\rangle$ & $\sigma$ & N &
               & $\langle$Age$\rangle$ & $\sigma$ & N & \\
\noalign{\smallskip}
\hline
\noalign{\smallskip}
`low-alpha' halo  & 11.14 & $\pm$0.42 &  5 &       
                  & 10.78 & $\pm$0.24 & 10 &      
                  & 10.46 & $\pm$0.32 & 10 &   
                  & $\cdots$ & $\cdots$ &  $\cdots$ &
                  & $\cdots$ & $\cdots$ &  $\cdots$ & \\ 
`high-alpha' halo & 11.70 & $\pm$0.51 &  4 &          
                  & 14.47 & $\pm$0.62 &  3 &      
                  & 12.66 & $\pm$0.67 &  2 &        
                  & 11.35 & $\pm$0.33 & 13 &
                  &  9.07 & $\pm$0.43 &  6 & \\
thick-disk        & 11.60 & $\pm$0.80 &  1 &        
                  & 12.01 & $\pm$0.54 &  6 &        
                  & 11.56 & $\pm$0.50 &  3 &         
                  & 10.46 & $\pm$0.58 &  3 &
                  &  $\cdots$ & $\cdots$ &  $\cdots$ & \\
`high-alpha' + thick-disk & 11.67 & $\pm$0.43 &  5 &
                          & 13.07 & $\pm$0.41 &  9 &
                          & 11.95 & $\pm$0.40 &  5 &
                          & 11.12 & $\pm$0.29 & 16 &
                          &  9.07 & $\pm$0.43 &  6 & \\

\noalign{\smallskip}
\hline
\end{tabular}

\end{table*}


\section{Orbital parameters}

\label{sect:Orbital Parameters}
The calculation of stellar space velocities is described in Paper I.
In summary, proper motions are acquired from the Tycho--2 catalogue 
(H{\o}g et al. \cite{hoeg00}) for the large majority of stars,
radial velocities from our own spectral analyses, and distances
from Hipparcos parallaxes (van Leeuwen \cite{leeuwen07}),
or from the photometric absolute magnitude calibration
by Schuster et al.  (\cite{schuster04}, \cite{schuster06}). 
The resulting values of the velocity components 
$U_{\rm LSR}, V_{\rm LSR}, W_{\rm LSR}$ with respect to the LSR
are given in Tables 3 and 4 in Paper I.  The typical errors of 
these components are $(\pm 12, \pm 16, \pm 9)$ km s$^{-1}$ 
with the principal contribution coming from the error in the distances.

\begin{figure}
\resizebox{\hsize}{!}{\includegraphics{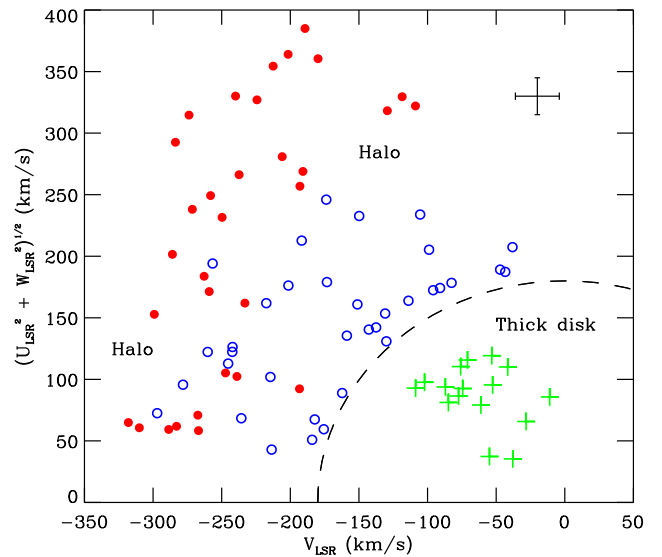}}
\caption{Toomre diagram for `high-$\alpha$' halo, `low-$\alpha$' halo,
and thick-disk stars with $\feh > -1.4$. The same symbols as in
Fig.~\ref{fig:logg-logteff1} are used. The long-dashed line corresponds to
$V_{\rm total} = 180\; \kmprs\ $.}
\label{fig:toomre}
\end{figure}

In Fig. \ref{fig:toomre} we have repeated the Toomre diagram from Paper I
for thick-disk and halo stars with $-1.4 < \feh < -0.4$. For this
range , it is possible to make a clear classification
into the `high-$\alpha$' and `low-$\alpha$' populations from the
\alphafe\---\feh\ diagram.  In the metallicity range 
$-1.6 < \feh < -1.4$, the two halo populations tend to merge in \alphafe,
and the classification is less clear.  For this reason, all following
figures with orbital parameters are confined to stars having
$\feh > -1.4$, i.e. 35 `high-$\alpha$', 32 `low-$\alpha$', and
16 thick-disk stars.


With these kinematic data, Galactic orbits of the stars
were computed backward in time 1, 2, and 5 Gyr in a detailed
semi-analytic model of the \object{Milky Way} potential, which has been
constrained to fit numerous observations of the Galactic bar and
spiral arms.  The Bulirsch-Stoer algorithm given by Press et al.\
(\cite{press92}) was used in the computations of the Galactic stellar
orbits.  Orbital parameters such as ${\rm r}_{\rm max}$,
${\rm z}_{\rm max}$, and ${\rm e}_{\rm max}$ have been derived, where
these are the maximum distances from the Galactic center and from the
Galactic disk, and the maximum orbital eccentricity obtained during
the orbital excursions of these integrations, respectively.  The orbital
integrations were carried out with both symmetrical and
non-symmetrical models.  The more complicated and more realistic
non-symmetrical model includes an axisymmetric background potential
as well as non-axisymmetric Galactic components:  bar and spiral
arms. The axisymmetric model used, with bulge, disk, and dark halo
components, is the Galactic model of Allen \& Santill\'an (\cite{AS91}),
scaled to give a rotational velocity of 254 \kms \ at the solar position,
a value recently found by Reid et al.~(\cite{RMZ09}).  The Sun-Galactic
center distance has been taken as R{$_0$} = 8.5 kpc, which is within the
range found by Reid et al.~(\cite{RMZ09}) (R$_0 = 8.6 \pm 0.6$ kpc).
In the Galactic potential the original scaled disk and bulge
components of the axisymmetric model are modified in order to
introduce the bar and spiral arms, keeping the same original scaled
mass.  All the mass in the original scaled spherical bulge is now used
to build the bar, and a fraction of the mass of the original scaled 
disk is employed to build the spiral arms.  Thus the remaining axisymmetric
components are just the diminished disk and the original scaled
spherical dark halo.

The Galactic three-dimensional potential for the spiral arms are
modeled following Pichardo et al.~(\cite{PMM03}). The spiral arms consist
of a superposition of inhomogeneous oblate spheroids, which can be
adjustable to better represent the available observations of the
Galactic spiral arms.  The spiral arms in the model trace the locus
found by Drimmel \& Spergel (\cite{DS01}), from K-band observations.  The
total mass of these arms is 3\% of the mass of the scaled axisymmetric
disk.  With this mass the mean ratio of the radial force due to the arms
to that of the axisymmetric background is around 10\%, in agreement
with the estimations by Patsis et al.~(\cite{PCG91}) for \object{Milky Way}-type
galaxies.  Also, the parameter Q$_t$, which is the ratio of the maximum
azimuthal force of the spiral arms at a given Galactocentric distance
on the Galactic plane, to the radial axisymmetric force at that distance
(Sanders \& Tubbs \cite{sanders80}; Combes \& Sanders \cite{combes81})
reaches a maximum value (Q$_t)_{max}$ = Q$_s = 0.12$, which is appropriate
for a Hubble-type galaxy like the \object{Milky Way} (Buta et al.\
\cite{buta04}; Buta et al. \cite{buta05}).  The self-consistency of the
spiral arms is tested through the reinforcement of the spiral potential
by the stellar orbits; see details in Patsis et al.~(\cite{PCG91}) and
Pichardo et al.~(\cite{PMM03}).  In our computations, 20 $\kmskpc$ has been
used for the pattern speed of the spiral arms (Martos et al.~\cite{MHY04}).

For the Galactic bar, the model of superposition of inhomogeneous 
ellipsoids given by Pichardo et al.~(\cite{PMM04}) (see their Appendix
C) has been utilized. This is a bar model which approximates the observed
boxy mass distribution of the Galactic bar, and is based on a model of
Freudenreich (\cite{F98}) of COBE/DIRBE observations of the Galactic
center.  At its present position, the major axis of the bar makes an 
angle of approximately 20$^\circ$ with the Sun-Galactic center line.
The angular velocity of the bar has been taken as 60 $\kmskpc$; see for
example Debattista et al.~(\cite{DGS02}).  With this bar angular velocity
and that of the spiral arms, the solar position is close to the bar 1:2
outer Lindblad resonance (OLR) and the spiral-arms' 1:4 resonance.

The observationally motivated parameters of the non-axisymmetric
Galactic components are summarized in Table \ref{tab.param}. 
The corresponding parameters of the (unscaled) axisymmetric
components can be found in Allen \& Santill\'an (\cite{AS91}).


\begin{table}
\caption[ ]{Parameters of the non-axisymmetric Galactic components
(Pichardo et al.~2003, 2004) used in the Galactic potential}
\vspace{-0.3cm}
\setlength{\tabcolsep}{0.01cm}
\begin{tabular}{lcc}
\noalign{\smallskip}
\hline\hline
\noalign{\smallskip}
Parameter &  Value  &   References \\
\noalign{\smallskip}
\hline
\noalign{\smallskip}
{\it Spiral Arms}\\
\noalign{\smallskip}
\hline
locus                             & Bisymmetric (Logthm)                 & 1 \\
pitch angle                       & $15.5^{\circ}$                       & 2 \\
external limit                    & 12 kpc                               & 2 \\
scale length                      & $2.5\ {\rm kpc}$                     & Disk based\\
force contrast                    & $\sim$ 10 \%                         & 3 \\
pattern speed                     & 20 km s$^{-1}$ kpc$^{-1}$             & 4 \\
\noalign{\smallskip}
\hline
\noalign{\smallskip}
{\it Bar}\\
\noalign{\smallskip}
\hline
major semi-axis                   &3.5 kpc                                 & 5 \\
scale lengths                     &1.7, 0.64, 0.44 kpc                     & 5 \\
present major axis angle with     &                                        &  \\
\,\,\, respect to the Sun-GC line & $20^{\circ}$                          & 6  \\
mass                              & $1.8 \times 10^{10}\ {\rm M}_\odot$  & 7 \\
angular velocity                  & 60 km s$^{-1}$ kpc$^{-1}$              & 8 \\

\noalign{\smallskip}
\hline

\label{tab.param}
\end{tabular}
\vspace{0.1cm}

References. --1)~Churchwell et al.~(\cite{CBM09}).  2)~Drimmel (\cite{D00}).  3)~Patsis et
al.~(\cite{PCG91}).  4)~Martos et al.~(\cite{MHY04}).  5)~Freudenreich (\cite{F98}).
6)~Gerhard (\cite{G02}).  7)~Calchi Novati et al.~(\cite{CLJ08}); Zhao (\cite{Z96}); Blum
(\cite{B95}); Dehnen \& Binney (\cite{DB98}); Dwek et al.~(\cite{DAH95}); Kent (\cite{K92}).
8)~ Debattista et al.~(\cite{DGS02}); Weiner \& Sellwood (\cite{WS99}); Fux (\cite{F99});
Ibata \& Gilmore (\cite{IG95}); Englmaier \& Gerhard (\cite{EG99}). \\
\
\end{table}

\begin{figure}
\resizebox{\hsize}{!}{\includegraphics{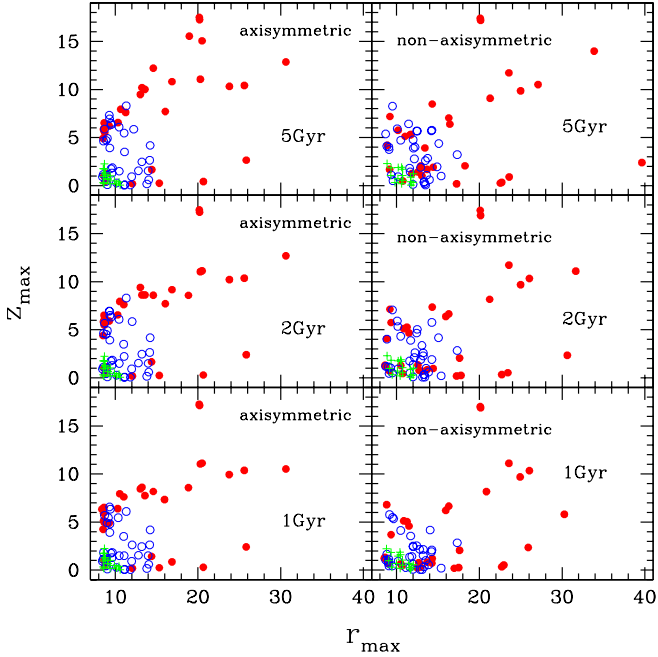}}
\caption{The `high-alpha' halo, `low-alpha' halo, and thick-disk
stars are plotted in the ${\rm z}_{\rm max}$ vs ${\rm r}_{\rm max}$ diagram for
orbital integration times of 1, 2, and 5 Gyr and for both the
axisymmetric and non-axisymmetric Galactic potentials.  The symbols
as in Fig.~\ref{fig:logg-logteff1}, and units of kpc for both axes.}
\label{fig:zmax-rmax6}
\end{figure}
\begin{figure}
\resizebox{\hsize}{!}{\includegraphics{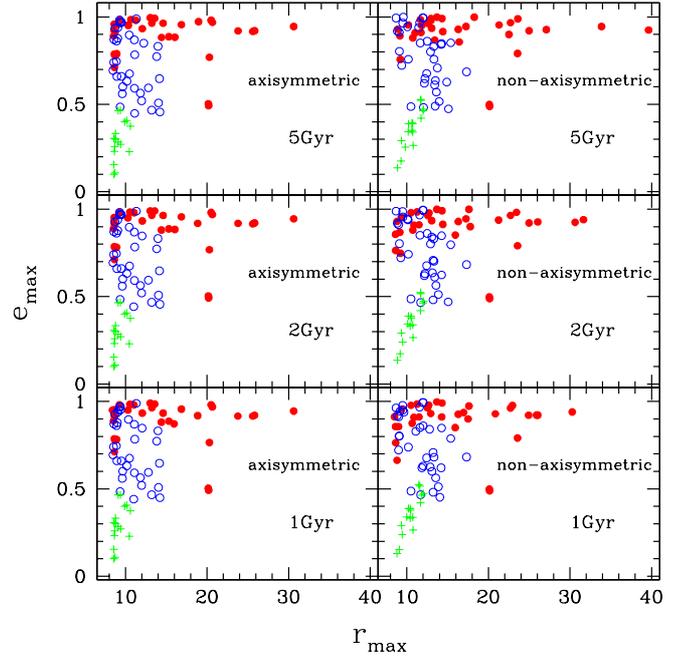}}
\caption{The `high-alpha' halo, `low-alpha' halo, and thick-disk
stars are plotted in the ${\rm e}_{\rm max}$ vs ${\rm r}_{\rm max}$ diagram for
orbital integration times of 1, 2, and 5 Gyr and for both the
axisymmetric and non-axisymmetric Galactic potentials.  The symbols
as in Fig.~\ref{fig:logg-logteff1}.}
\label{fig:e1max-rmax6}
\end{figure}

Figures~\ref{fig:zmax-rmax6} and \ref{fig:e1max-rmax6} show plots of ${\rm z}_{\rm max}$ and
${\rm e}_{\rm max}$ versus ${\rm r}_{\rm max}$, respectively, for the `high-' and `low-alpha' halo
stars, plus the thick-disk stars.  (${\rm z}_{\rm max}$ is the extreme maximum of $|$z$|$.)   In
the left halves of these figures the orbital parameters are plotted from the axisymmetric Galactic
model for the integration times of 5, 2, and 1 Gyr, from top to bottom, and in the right halves,
the same from the non-axisymmetric model.  In Figs.~\ref{fig:mgfe-e1max.2} and \ref{fig:nafe-e1max.2}
are plotted the \mgfe\ and \nafe\  abundances as a function of the orbital parameter ${\rm e}_{\rm max}$,
respectively, again for these same three Galactic stellar components.  Figs.~\ref{fig:mgfe-rmax.2}
and \ref{fig:nafe-rmax.2} plot these same abundances against ${\rm r}_{\rm max}$.  In these last four
figures the upper panels make use of orbital parameters from integrations with the axisymmetric
Galactic model, and from the non-axisymmetric model in the lower panels.  In all of these figures
the symbols are the same as in Fig.~\ref{fig:logg-logteff1}.

Table~4 (online) gives the stellar ages and orbital parameters derived in Sections \ref{sect:ages}
and  \ref{sect:Orbital Parameters} for the halo and thick-disk stars of this paper.  Columns 1 and
2 give the stellar identifications and the stellar ages with estimated errors, respectively.
Columns 3--9 give the extreme minimum and maximum values, over 5 Gyr, for r$_{\rm min}$, 
r$_{\rm max}$, $|$z$|_{\rm max}$, e$_{\rm min}$, e$_{\rm max}$, h$_{\rm min}$, and h$_{\rm max}$,
respectively (where h is the angular momentum per unit mass of the star).  For each star the first
line gives the orbital parameters for the non-axisymmetric case and the second line for the
axisymmetric.  The extreme maximum values of r, $|$z$|$, and e are those plotted in
Figs.~\ref{fig:zmax-rmax6}--\ref{fig:nafe-rmax.2}.


\onltab{4}{
\onecolumn
\begin{longtable}{lccccccccccc}
\caption[ ]{Stellar ages, and orbital parameters with the non-axisymmetric and axisymmetric potentials
(5 Gyr integrations).} \\ 
\noalign{\smallskip}
\hline\hline
\noalign{\smallskip}
Star & Age$\pm {\sigma}$Age & r$_{\rm min}$ & r$_{\rm max}$ & $|$z$|_{\rm max}$ & e$_{\rm min}$ & e$_{\rm max}$ & h$_{\rm min}$ & h$_{\rm max}$ \\
     & (Gyr)                 & (kpc)          & (kpc)         & (kpc)             &  &  &(/10 {\rm km\,s$^{-1}$kpc})   & (/10 {\rm km\,s$^{-1}$kpc}) \\
\noalign{\smallskip}
\hline
\noalign{\smallskip}
\noalign{\smallskip}
\endfirsthead
\caption{continued} \\
\hline\hline
\noalign{\smallskip}
Star & Age$\pm {\sigma}$Age & r$_{\rm min}$ & r$_{\rm max}$ & $|$z$|_{\rm max}$ & e$_{\rm min}$ & e$_{\rm max}$ & h$_{\rm min}$ & h$_{\rm max}$ \\
 & (Gyr) & (kpc) & (kpc) & (kpc) &  &  &(/10 {\rm km\,s$^{-1}$kpc}) & (/10 {\rm km\,s$^{-1}$kpc}) \\
\noalign{\smallskip}
\hline
\noalign{\smallskip}
\noalign{\smallskip}
\endhead
\hline
\endfoot
BD-21 3420 & 13.6 $\pm$ 1.5 & 4.48 & 10.62 & ~1.90 & 0.236 &  0.393 & $+137.82$ & $+164.79$   \\
           &                & 4.54 & ~8.57 & ~1.76 & 0.299 &  0.306 & $+141.71$ & $+141.71$   \\
CD-33 3337 & 11.6 $\pm$ 0.8 & 6.73 & ~8.85 & ~2.30 & 0.061 &  0.136 & $+176.20$ & $+184.23$   \\
           &                & 6.97 & ~8.71 & ~2.25 & 0.105 &  0.108 & $+181.26$ & $+181.26$   \\
CD-43 6810 & ~5.3 $\pm$ 1.0 & 1.96 & 12.43 & ~5.69 & 0.375 &  0.677 & ~$+70.25$ & $+112.54$   \\
           &                & 1.93 & ~9.50 & ~1.76 & 0.641 &  0.660 & ~$+79.33$ & ~$+79.33$   \\
CD-45 3283 & ...            & 1.00 & 24.99 & ~9.86 & 0.869 &  0.921 & ~$+22.19$ & ~$+46.83$   \\
           &                & 0.99 & 23.80 & 10.32 & 0.866 &  0.920 & ~$+39.13$ & ~$+39.13$   \\
CD-51 4628 & ~6.9 $\pm$ 1.4 & 2.63 & 23.57 & 11.73 & 0.671 &  0.791 & ~$+92.74$ & $+118.24$   \\
           &                & 2.65 & 20.29 & 11.06 & 0.707 &  0.769 & $+101.96$ & $+101.96$   \\
CD-57 1633 & 10.5 $\pm$ 1.6 & 0.09 & 23.61 & ~0.90 & 0.961 &  0.989 & ~$-13.12$ & ~$+28.43$   \\
           &                & 0.20 & 20.50 & 15.06 & 0.794 &  0.981 & ~$+15.06$ & ~$+15.06$   \\
CD-61 282  & 14.8 $\pm$ 1.9 & 0.48 & 17.25 & ~0.19 & 0.843 &  0.945 & ~$-59.24$ & ~$-23.26$   \\
           &                & 0.38 & 16.86 & 10.81 & 0.716 &  0.956 & ~$-26.76$ & ~$-26.76$   \\
G05-19     & 13.4 $\pm$ 1.4 & 0.45 & 21.31 & ~9.08 & 0.847 &  0.954 & ~$-20.16$ & ~~$-3.93$   \\
           &                & 0.25 & 18.96 & 15.54 & 0.808 &  0.973 & ~$-16.25$ & ~$-16.25$   \\
G05-36     & 10.3 $\pm$ 1.0 & 0.19 & 11.57 & ~1.82 & 0.744 &  0.967 & ~$-49.96$ & ~~$-8.60$   \\
           &                & 0.66 & ~8.91 & ~4.71 & 0.567 &  0.862 & ~$-37.30$ & ~$-37.30$   \\
G05-40     & 12.7 $\pm$ 0.7 & 0.80 & 12.50 & ~5.67 & 0.654 &  0.861 & ~$+18.24$ & ~$+56.77$   \\
           &                & 0.56 & 10.47 & ~6.09 & 0.615 &  0.897 & ~$+32.01$ & ~$+32.01$   \\
G13-38     & ...            & 2.38 & 12.22 & ~4.01 & 0.505 &  0.647 & ~$+85.21$ & $+119.55$   \\
           &                & 2.39 & ~9.67 & ~1.86 & 0.586 &  0.601 & ~$+94.05$ & ~$+94.05$   \\
G15-23     & ...            & 1.14 & 10.19 & ~0.93 & 0.650 &  0.758 & ~$+47.59$ & ~$+96.69$   \\
           &                & 1.53 & ~8.47 & ~0.91 & 0.682 &  0.694 & ~$+66.23$ & ~$+66.23$   \\
G16-20     & ...            & 1.27 & 15.39 & ~2.10 & 0.678 &  0.779 & ~$+55.82$ & $+124.48$   \\
           &                & 2.07 & 12.67 & ~1.82 & 0.705 &  0.719 & ~$+93.49$ & ~$+93.49$   \\
G18-28     & ...            & 0.35 & ~8.83 & ~4.11 & 0.585 &  0.919 & ~$+16.04$ & ~$+52.51$   \\
           &                & 0.59 & ~8.59 & ~4.64 & 0.554 &  0.871 & ~$+34.19$ & ~$+34.19$   \\
G18-39     & 11.4 $\pm$ 1.0 & 0.03 & 12.00 & ~4.77 & 0.701 &  0.994 &  $-34.64$ & ~~$+2.37$   \\
           &                & 0.06 & 11.34 & ~8.30 & 0.734 &  0.989 & ~~$-4.66$ & ~~$-4.66$   \\
G20-15     & ...            & 5.58 & 16.90 & ~9.91 & 0.441 &  0.495 & $+158.55$ & $+165.94$   \\
           &                & 5.76 & 16.46 & ~9.68 & 0.448 &  0.480 & $+163.07$ & $+163.07$   \\
G21-22     & 12.5 $\pm$ 1.2 & 0.06 & 14.31 & ~8.48 & 0.702 &  0.991 & ~$-12.23$ & ~~$+8.35$   \\
           &                & 0.05 & 13.59 & 10.01 & 0.732 &  0.992 & ~~$-1.37$ & ~~$-1.37$   \\
G24-13     & 11.3 $\pm$ 2.0 & 5.25 & 15.10 & ~4.38 & 0.374 &  0.474 & $+173.18$ & $+190.55$   \\
           &                & 5.30 & 14.23 & ~4.18 & 0.443 &  0.456 & $+183.30$ & $+183.30$   \\
G24-25     & ...            & 0.19 & 12.25 & ~8.24 & 0.586 &  0.963 & ~$-19.19$ & ~$+18.45$   \\
           &                & 0.14 & 11.54 & ~7.25 & 0.699 &  0.976 & ~~$+8.93$ & ~~$+8.93$   \\
G31-55     & 17.2 $\pm$ 2.1 & 0.93 & 12.91 & ~1.89 & 0.644 &  0.801 & ~$+40.45$ & $+107.95$   \\
           &                & 1.28 & ~8.57 & ~1.17 & 0.704 &  0.740 & ~$+59.47$ & ~$+59.47$   \\
G46-31     & 10.2 $\pm$ 1.1 & 0.30 & 11.84 & ~1.25 & 0.588 &  0.951 & ~$-70.73$ & ~$-15.94$   \\
           &                & 1.03 & ~8.63 & ~1.01 & 0.773 &  0.786 & ~$-48.39$ & ~$-48.39$   \\
G49-19     & ~9.2 $\pm$ 1.3 & 0.02 & ~8.68 & ~5.37 & 0.529 &  0.994 & ~$-14.66$ & ~$+16.24$   \\
           &                & 0.28 & ~8.72 & ~5.88 & 0.604 &  0.936 & ~$+15.61$ & ~$+15.61$   \\
G53-41     & 12.4 $\pm$ 1.2 & 1.22 & ~9.20 & ~7.19 & 0.454 &  0.757 & ~$-51.30$ & ~$-32.37$   \\
           &                & 1.43 & ~8.66 & ~6.52 & 0.485 &  0.712 & ~$-38.21$ & ~$-38.21$   \\
G56-30     & 11.8 $\pm$ 1.4 & 0.53 & ~9.15 & ~1.68 & 0.754 &  0.891 & ~$-42.70$ & ~$-18.39$   \\
           &                & 0.40 & ~8.73 & ~5.50 & 0.581 &  0.913 & ~$-24.82$ & ~$-24.82$   \\
G56-36     & ~8.7 $\pm$ 1.4 & 0.18 & 12.58 & ~1.43 & 0.855 &  0.963 & ~~$-0.57$ & ~$+45.59$   \\
           &                & 0.09 & ~9.27 & ~6.20 & 0.661 &  0.981 & ~~$+6.20$ & ~~$+6.20$   \\
G57-07     & ~9.9 $\pm$ 1.4 & 1.45 & ~9.37 & ~1.11 & 0.607 &  0.723 & ~$+60.41$ & ~$+98.45$   \\
           &                & 1.83 & ~8.95 & ~1.04 & 0.637 &  0.660 & ~$+78.54$ & ~$+78.54$   \\
G63-26     & ...            & 1.01 & 10.68 & ~1.23 & 0.618 &  0.826 & ~$-80.45$ & ~$-45.17$   \\
           &                & 1.47 & ~9.07 & ~1.03 & 0.702 &  0.721 & ~$-65.94$ & ~$-65.94$   \\
G66-22     & ...            & 0.85 & 22.52 & ~0.26 & 0.843 &  0.900 & ~$+46.19$ & ~$+98.79$   \\
           &                & 0.91 & 15.33 & ~0.25 & 0.887 &  0.887 & ~$+53.54$ & ~$+53.54$   \\
\\
G74-32     & ~7.0 $\pm$ 2.3 & 1.77 & 13.75 & ~2.85 & 0.641 &  0.708 & ~$+68.86$ & $+115.26$   \\
           &                & 2.03 & 10.50 & ~1.50 & 0.660 &  0.675 & ~$+87.58$ & ~$+87.58$   \\
G75-31     & 10.7 $\pm$ 0.5 & 1.17 & 16.43 & ~6.39 & 0.797 &  0.857 & ~$+26.71$ & ~$+50.89$   \\
           &                & 0.99 & 16.06 & ~7.71 & 0.776 &  0.884 & ~$+40.86$ & ~$+40.86$   \\
G81-02     & 10.3 $\pm$ 0.6 & 0.88 & 11.65 & ~1.43 & 0.794 &  0.851 & ~$+37.32$ & ~$+62.39$   \\
           &                & 0.78 & 11.09 & ~5.46 & 0.641 &  0.869 & ~$+44.44$ & ~$+44.44$   \\
G82-05     & ...            & 1.05 & 39.64 & ~2.39 & 0.906 &  0.925 & ~$+52.28$ & $+109.59$   \\
           &                & 1.07 & 25.84 & ~2.65 & 0.910 &  0.921 & ~$+64.46$ & ~$+64.46$   \\
G85-13     & ~6.8 $\pm$ 2.7 & 4.44 & 13.97 & ~0.61 & 0.469 &  0.516 & $+171.43$ & $+182.88$   \\
           &                & 4.58 & 14.05 & ~0.60 & 0.507 &  0.508 & $+180.97$ & $+180.97$   \\
G87-13     & ~9.7 $\pm$ 0.5 & 0.42 & 13.05 & ~1.02 & 0.891 &  0.935 & ~$-40.03$ & ~$-17.62$   \\
           &                & 0.42 & 12.07 & ~0.20 & 0.932 &  0.933 & ~$-27.76$ & ~$-27.76$   \\
G94-49     & ...            & 2.65 & 13.41 & ~0.17 & 0.518 &  0.611 & $+103.05$ & $+148.12$   \\
           &                & 3.30 & 11.84 & ~0.06 & 0.564 &  0.564 & $+136.01$ & $+136.01$   \\
G96-20     & ...            & 4.05 & 12.04 & ~3.89 & 0.393 &  0.482 & $+138.05$ & $+148.69$   \\
           &                & 4.24 & 11.13 & ~3.50 & 0.424 &  0.448 & $+140.08$ & $+140.08$   \\
G98-53     & 11.5 $\pm$ 1.1 & 0.50 & 11.02 & ~5.12 & 0.656 &  0.909 & ~$-32.68$ & ~~$-1.82$   \\
           &                & 0.11 & 11.27 & ~7.62 & 0.685 &  0.981 & ~~$-7.69$ & ~~$-7.69$   \\
G99-21     & 10.6 $\pm$ 3.3 & 0.96 & ~9.09 & ~3.75 & 0.516 &  0.803 & ~$+43.17$ & ~$+79.31$   \\
           &                & 1.30 & ~8.89 & ~0.43 & 0.740 &  0.746 & ~$+61.88$ & ~$+61.88$   \\
G112-43    & 11.4 $\pm$ 0.5 & 6.73 & 20.11 & 17.42 & 0.451 &  0.499 & $+114.74$ & $+116.87$   \\
           &                & 6.64 & 20.16 & 17.51 & 0.460 &  0.503 & $+115.49$ & $+115.49$   \\
G112-44    & 15.7 $\pm$ 1.5 & 6.94 & 20.15 & 17.20 & 0.441 &  0.488 & $+123.16$ & $+125.21$   \\
           &                & 6.85 & 20.20 & 17.26 & 0.452 &  0.492 & $+123.98$ & $+123.98$   \\
G114-42    & 13.7 $\pm$ 2.8 & 0.90 & 33.87 & 13.99 & 0.885 &  0.945 & ~$+43.52$ & ~$+65.40$   \\
           &                & 0.86 & 30.63 & 12.86 & 0.883 &  0.945 & ~$+47.76$ & ~$+47.76$   \\
G119-64    & ...            & 0.98 & 15.12 & ~5.52 & 0.783 &  0.877 & ~$-39.24$ & ~$-25.24$   \\
           &                & 0.58 & 14.80 & ~8.33 & 0.736 &  0.925 & ~$-31.42$ & ~$-31.42$   \\
G121-12    & 10.5 $\pm$ 1.1 & 0.94 & 27.10 & 10.51 & 0.864 &  0.927 & ~$+35.80$ & ~$+54.60$   \\
           &                & 1.11 & 25.61 & 10.42 & 0.869 &  0.917 & ~$+45.55$ & ~$+45.55$   \\
G125-13    & ...            & 0.31 & 14.39 & 11.13 & 0.635 &  0.946 & ~~$-3.96$ & ~$+33.53$   \\
           &                & 0.36 & 14.73 & ~9.49 & 0.703 &  0.951 & ~$+24.00$ & ~$+24.00$   \\
G127-26    & 10.3 $\pm$ 0.6 & 3.70 & 13.63 & ~1.09 & 0.515 &  0.570 & $+148.08$ & $+162.91$   \\
           &                & 3.79 & 11.99 & ~0.93 & 0.518 &  0.520 & $+148.83$ & $+148.83$   \\
G150-40    & 11.2 $\pm$ 1.0 & 0.69 & 10.74 & ~0.48 & 0.698 &  0.880 & ~$-66.11$ & ~$-32.75$   \\
           &                & 1.10 & ~8.75 & ~0.40 & 0.774 &  0.777 & ~$-54.35$ & ~$-54.35$   \\
G159-50    & 12.2 $\pm$ 2.5 & 1.06 & 14.20 & ~5.68 & 0.672 &  0.842 & ~$+45.08$ & ~$+78.24$   \\
           &                & 0.99 & 12.24 & ~5.86 & 0.681 &  0.850 & ~$+52.97$ & ~$+52.97$   \\
G161-73    & 10.7 $\pm$ 0.6 & 0.01 & 13.68 & ~1.68 & 0.909 &  0.999 & ~$-22.93$ & ~~$+7.37$   \\
           &                & 0.03 & 13.04 & ~9.48 & 0.701 &  0.995 & ~~$-0.02$ & ~~$-0.02$   \\
G170-56    & 10.3 $\pm$ 0.6 & 0.13 & 12.94 & ~1.87 & 0.792 &  0.979 & ~$-40.88$ & ~$+13.65$   \\
           &                & 0.49 & ~8.54 & ~4.91 & 0.569 &  0.891 & ~$-29.53$ & ~$-29.53$   \\
G172-61    & ...            & 0.31 & ~8.89 & ~4.19 & 0.637 &  0.930 & ~$-28.44$ & ~~$-3.52$   \\
           &                & 0.33 & ~8.60 & ~5.91 & 0.777 &  0.926 & ~$-11.12$ & ~$-11.12$   \\
G176-53    & ...            & 0.01 & 18.27 & ~2.06 & 0.933 &  0.999 & ~$-24.29$ & ~$+24.66$   \\
           &                & 0.23 & 13.24 & 10.19 & 0.684 &  0.965 & ~$-15.25$ & ~$-15.25$   \\
G180-24    & 12.3 $\pm$ 1.1 & 0.00 & ~9.57 & ~0.51 & 0.942 &  0.999 & ~$-12.29$ & ~$+11.90$   \\
           &                & 0.13 & ~9.32 & ~6.95 & 0.561 &  0.971 & ~~$+8.22$ & ~~$+8.22$   \\
G187-18    & ~7.9 $\pm$ 2.9 & 2.58 & 12.15 & ~2.78 & 0.484 &  0.620 & ~$+96.88$ & $+127.86$   \\
           &                & 2.71 & ~9.62 & ~1.30 & 0.550 &  0.560 & $+106.09$ & $+106.09$   \\
G188-22    & 12.9 $\pm$ 1.0 & 3.25 & 14.90 & ~1.96 & 0.575 &  0.627 & $+131.26$ & $+159.73$   \\
           &                & 3.26 & 12.81 & ~1.55 & 0.588 &  0.594 & $+134.15$ & $+134.15$   \\
G192-43    & ~7.4 $\pm$ 1.4 & 0.32 & 22.71 & ~0.34 & 0.947 &  0.967 & ~$+14.29$ & ~$+36.32$   \\
           &                & 0.32 & 20.66 & ~0.42 & 0.968 &  0.969 & ~$+23.82$ & ~$+23.82$   \\
G232-18    & ...            & 0.25 & 10.09 & ~5.92 & 0.658 &  0.944 & ~~$-3.39$ & ~$+29.21$   \\
           &                & 0.15 & ~9.51 & ~6.33 & 0.639 &  0.968 & ~$+10.66$ & ~$+10.66$   \\
HD3567     & 11.0 $\pm$ 0.5 & 0.08 & 11.66 & ~5.28 & 0.672 &  0.985 & ~$-19.34$ & ~$+16.54$   \\
           &                & 0.08 & 10.67 & ~7.94 & 0.663 &  0.985 & ~~$-3.71$ & ~~$-3.71$   \\
\\
HD17820    & 11.2 $\pm$ 0.9 & 4.39 & 10.21 & ~1.87 & 0.299 &  0.388 & $+143.13$ & $+158.37$   \\
           &                & 4.40 & ~8.79 & ~1.15 & 0.327 &  0.333 & $+144.64$ & $+144.64$   \\
HD22879    & 12.5 $\pm$ 1.2 & 4.12 & 12.04 & ~1.24 & 0.415 &  0.473 & $+152.07$ & $+171.06$   \\
           &                & 4.24 & ~9.90 & ~0.48 & 0.399 &  0.400 & $+151.77$ & $+151.77$   \\
HD25704    & ~9.8 $\pm$ 1.3 & 4.68 & 11.74 & ~0.06 & 0.358 &  0.420 & $+171.44$ & $+189.87$   \\
           &                & 4.83 & 10.61 & ~0.05 & 0.375 &  0.375 & $+171.12$ & $+171.12$   \\
HD51754    & ~8.5 $\pm$ 1.7 & 0.82 & 15.39 & ~1.00 & 0.757 &  0.852 & ~$+39.83$ & $+107.67$   \\
           &                & 1.77 & 13.83 & ~0.17 & 0.772 &  0.773 & ~$+88.23$ & ~$+88.23$   \\
HD59329    & ...            & 0.61 & 11.03 & ~0.69 & 0.671 &  0.895 & ~$-69.69$ & ~$-29.29$   \\
           &                & 0.86 & ~9.51 & ~0.54 & 0.829 &  0.833 & ~$-46.34$ & ~$-46.34$   \\
HD76932    & 11.7 $\pm$ 0.6 & 4.47 & 10.73 & ~1.70 & 0.241 &  0.389 & $+138.10$ & $+167.22$   \\
           &                & 4.72 & ~8.77 & ~1.31 & 0.295 &  0.300 & $+150.30$ & $+150.30$   \\
HD97320    & 10.8 $\pm$ 0.8 & 6.19 & 10.78 & ~0.55 & 0.220 &  0.265 & $+197.90$ & $+207.49$   \\
           &                & 6.57 & 10.47 & ~0.38 & 0.229 &  0.229 & $+205.64$ & $+205.64$   \\
HD103723   & ~9.9 $\pm$ 0.7 & 0.59 & 13.42 & ~3.91 & 0.564 &  0.868 & ~$+27.62$ & ~$+96.71$   \\
           &                & 1.06 & ~8.94 & ~1.21 & 0.754 &  0.789 & ~$+52.13$ & ~$+52.13$   \\
HD105004   & 11.6 $\pm$ 1.8 & 0.21 & 10.18 & ~5.75 & 0.652 &  0.954 & ~~$+2.02$ & ~$+24.19$   \\
           &                & 0.19 & ~9.10 & ~6.33 & 0.608 &  0.958 & ~$+13.21$ & ~$+13.21$   \\
HD106516   & ...            & 4.83 & 10.15 & ~0.72 & 0.230 &  0.343 & $+156.44$ & $+176.40$   \\
           &                & 5.07 & ~9.11 & ~0.69 & 0.283 &  0.285 & $+163.81$ & $+163.81$   \\
HD111980   & 14.6 $\pm$ 1.1 & 0.86 & 14.26 & ~5.77 & 0.608 &  0.848 & ~$+29.72$ & ~$+79.14$   \\
           &                & 1.28 & 13.99 & ~1.54 & 0.812 &  0.832 & ~$+67.08$ & ~$+67.08$   \\
HD113679   & 13.1 $\pm$ 0.8 & 0.03 & 11.98 & ~0.13 & 0.921 &  0.995 & ~$-22.55$ & ~$+12.40$   \\
           &                & 0.29 & ~9.06 & ~0.10 & 0.938 &  0.938 & ~$-19.40$ & ~$-19.40$   \\
HD114762A  & 10.1 $\pm$ 0.8 & 5.10 & 10.71 & ~1.05 & 0.139 &  0.341 & $+155.46$ & $+184.33$   \\
           &                & 5.40 & ~9.41 & ~1.02 & 0.267 &  0.271 & $+171.12$ & $+171.12$   \\
HD120559   & ...            & 5.70 & ~9.80 & ~0.32 & 0.079 &  0.255 & $+171.20$ & $+189.88$   \\
           &                & 6.28 & ~8.57 & ~0.33 & 0.153 &  0.154 & $+183.36$ & $+183.36$   \\
HD121004   & ~8.2 $\pm$ 2.7 & 0.28 & 11.45 & ~6.42 & 0.665 &  0.939 & ~~$-3.15$ & ~$+40.96$   \\
           &                & 0.14 & ~9.49 & ~6.50 & 0.623 &  0.971 & ~~$+9.83$ & ~~$+9.83$   \\
HD126681   & ...            & 6.32 & ~9.29 & ~1.04 & 0.085 &  0.176 & $+179.07$ & $+194.96$   \\
           &                & 7.01 & ~8.57 & ~0.94 & 0.098 &  0.100 & $+191.24$ & $+191.24$   \\
HD132475   & ...            & 1.92 & ~9.74 & ~1.13 & 0.561 &  0.671 & $-100.58$ & ~$-81.39$   \\
           &                & 2.56 & ~8.60 & ~0.86 & 0.535 &  0.541 & ~$-98.37$ & ~$-98.37$   \\
HD148816   & 11.7 $\pm$ 0.6 & 0.14 & ~9.51 & ~8.26 & 0.425 &  0.965 & ~$-17.42$ & ~$+18.60$   \\
           &                & 0.08 & ~9.29 & ~7.33 & 0.570 &  0.982 & ~~$-5.14$ & ~~$-5.14$   \\
HD159482   & ~9.8 $\pm$ 1.9 & 4.61 & 13.48 & ~2.60 & 0.420 &  0.490 & $+165.71$ & $+177.35$   \\
           &                & 4.80 & 13.22 & ~2.47 & 0.461 &  0.467 & $+174.57$ & $+174.57$   \\
HD160693   & ~9.2 $\pm$ 1.5 & 2.88 & 17.33 & ~3.22 & 0.626 &  0.686 & $+120.00$ & $+147.54$   \\
           &                & 3.03 & 14.17 & ~2.64 & 0.637 &  0.647 & $+126.93$ & $+126.93$   \\
HD163810   & ...            & 0.57 & 14.42 & ~1.91 & 0.854 &  0.916 & ~$+29.97$ & ~$+60.56$   \\
           &                & 0.90 & 14.42 & ~1.66 & 0.860 &  0.883 & ~$+51.93$ & ~$+51.93$   \\
HD175179   & ~8.7 $\pm$ 2.2 & 3.34 & 11.72 & ~0.41 & 0.444 &  0.528 & $+123.65$ & $+153.54$   \\
           &                & 3.41 & ~9.37 & ~0.20 & 0.466 &  0.466 & $+128.45$ & $+128.45$   \\
HD177095   & 15.6 $\pm$ 3.6 & 1.14 & 13.33 & ~0.39 & 0.624 &  0.758 & ~$+48.93$ & $+125.28$   \\
           &                & 2.29 & 10.17 & ~0.28 & 0.632 &  0.632 & ~$+98.58$ & ~$+98.58$   \\
HD179626   & 14.0 $\pm$ 0.8 & 0.42 & ~9.07 & ~4.00 & 0.584 &  0.912 & ~~$+5.97$ & ~$+47.52$   \\
           &                & 0.59 & ~9.06 & ~4.92 & 0.569 &  0.878 & ~$+33.75$ & ~$+33.75$   \\
HD189558   & 13.8 $\pm$ 1.7 & 3.34 & 11.70 & ~0.89 & 0.441 &  0.523 & $+121.83$ & $+151.56$   \\
           &                & 3.32 & ~9.09 & ~0.66 & 0.463 &  0.465 & $+123.01$ & $+123.01$   \\
HD193901   & 14.7 $\pm$ 2.6 & 0.48 & 11.74 & ~5.31 & 0.683 &  0.911 & ~~$+5.78$ & ~$+42.46$   \\
           &                & 0.25 & 10.35 & ~6.55 & 0.687 &  0.952 & ~$+17.50$ & ~$+17.50$   \\
HD194598   & ~8.6 $\pm$ 1.8 & 0.05 & 12.94 & ~2.04 & 0.883 &  0.990 & ~$-20.72$ & ~$+36.10$   \\
           &                & 0.18 & ~8.61 & ~5.79 & 0.621 &  0.959 & ~$-11.66$ & ~$-11.66$   \\
HD199289   & 12.2 $\pm$ 1.5 & 4.86 & 10.61 & ~0.21 & 0.201 &  0.350 & $+157.14$ & $+182.72$   \\
           &                & 5.39 & ~8.64 & ~0.19 & 0.231 &  0.231 & $+168.46$ & $+168.46$   \\
HD205650   & 15.3 $\pm$ 2.1 & 3.91 & 12.00 & ~0.32 & 0.321 &  0.463 & $+133.80$ & $+173.13$   \\
           &                & 4.28 & 10.15 & ~0.21 & 0.407 &  0.407 & $+154.85$ & $+154.85$   \\
\\
HD219617   & ...            & 1.29 & 20.52 & ~0.75 & 0.787 &  0.833 & ~$+63.21$ & $+123.39$   \\
           &                & 1.32 & 13.21 & ~0.63 & 0.816 &  0.819 & ~$+68.64$ & ~$+68.64$   \\
HD222766   & 20.9 $\pm$ 1.9 & 0.98 & 13.31 & ~0.33 & 0.752 &  0.831 & ~$+46.78$ & ~$+89.99$   \\
           &                & 1.35 & 11.15 & ~0.09 & 0.783 &  0.784 & ~$+67.92$ & ~$+67.92$   \\
HD230409   & ...            & 2.22 & 13.50 & ~0.06 & 0.564 &  0.636 & ~$+90.28$ & $+141.32$   \\
           &                & 2.84 & 11.10 & ~0.04 & 0.592 &  0.592 & $+119.38$ & $+119.38$   \\
HD233511   & ...            & 0.27 & 11.69 & ~0.67 & 0.895 &  0.948 & ~~$+5.66$ & ~$+33.55$   \\
           &                & 0.14 & ~9.76 & ~6.78 & 0.652 &  0.970 & ~~$+9.19$ & ~~$+9.19$   \\
HD237822   & ~8.0 $\pm$ 2.2 & 3.20 & 10.50 & ~4.71 & 0.376 &  0.487 & ~$+97.75$ & $+114.41$   \\
           &                & 3.24 & ~9.35 & ~3.92 & 0.430 &  0.485 & $+105.76$ & $+105.76$   \\
HD241253   & 11.3 $\pm$ 1.6 & 5.08 & ~9.34 & ~1.86 & 0.233 &  0.292 & $+151.95$ & $+160.14$   \\
           &                & 5.10 & ~8.69 & ~1.74 & 0.253 &  0.259 & $+154.14$ & $+154.14$   \\
HD250792A  & ...            & 0.58 & 16.31 & ~7.04 & 0.775 &  0.929 & ~~$+9.80$ & ~$+34.85$   \\
           &                & 0.26 & 14.61 & 12.22 & 0.708 &  0.964 & ~$+14.06$ & ~$+14.06$   \\
HD284248   & ...            & 1.53 & 33.04 & ~3.08 & 0.876 &  0.900 & ~$+76.56$ & $+117.04$   \\
           &                & 1.59 & 26.20 & ~2.81 & 0.878 &  0.886 & ~$+88.96$ & ~$+88.96$   \\

\noalign{\smallskip}
\hline
\end{longtable}
}


\section{Discussion and conclusions}

The metallicity trends of Papers I and II can be
explained from existing nucleosynthesis simulations if the `high-alpha' halo stars
formed in regions with such a high star formation rate (SFR) that only massive stars and
Type II supernovae contributed to the chemical enrichment, while the `low-alpha'
halo stars, on the other hand, originated from systems with a slower chemical
evolution, characterized by additional enrichment from Type Ia supernovae and
low-mass AGB stars.  The more massive sub-units of the Galactic formation had higher
densities, higher SFRs, more rapid chemical evolutions, and therefore reached higher
metallicities (\feh\ $\sim -0.4$ dex) before the ignition of the first SNeIa (see the
\alphafe\ versus \feh\ diagrams in Fig.~1 of McWilliam \cite{McWilliam97}, or Fig.~1 of
Zolotov et al.\ \cite{zolotov10}) leading to the `high-alpha' halo stars, while the smaller
sub-units had smaller densities, lower SFRs, slower chemical evolution, and therefore lower
metallicities at the ``knee'' (\feh\ $\la -1.5$ dex, in this \alphafe\ versus \feh\ diagram)
corresponding to the ignition of the SNeIa, producing `low-alpha' halo stars at the
higher metallicities.  The question then becomes: which of the many scenarios presented
in the literature to explain halo components best fits these observed abundance patterns,
relative-age differences, stellar kinematics, and orbital parameters of these halo stars?
For example, does the classic ELS-plus-SZ duality with a rapid collapsed, more centrally
distributed halo component versus a more exterior component originating from
``protogalactic fragments'' accreted over several Gyr?  Or, does the
scattering of ``$in \: situ$'' stars from a primeval bulge or disk into the inner halo versus
the accretion of stars from ``subhalos'' $\grave{a}\ la$ Zolotov et al.\ (\cite{zolotov09},
\cite{zolotov10}) or Purcell et al.\ (\cite{purcell10})?  Or, does the early accretion and
merging of massive satellites but with a dichotomy of the mass distribution via the models
of Robertson et al.\ (\cite{Robertson05}), Bullock \& Johnston (\cite{Bullock05}), and Font
et al.\ (\cite{font06a}, \cite{font06b}) explain best these two halo components detected and
studied here?

Our age determinations for the `high-' and `low-alpha' halo stars are in agreement with
the conclusions of Hammer et al.\ (\cite{hammer07}) and Puech et al.\ (\cite{puech08}) that
the \object{Milky Way} appears to have been an ``exceptionally quiet galaxy,'' compared to
\object{M31} for example, having escaped any major mergers or accretions for the last
$\sim 10$ Gyr.  Nearly all of the stellar ages, or average ages, in Figs.~\ref{fig:logg-logteff1}
and \ref{fig:logg-logteff2} and in Table~\ref{table:ave_ages} are greater than 9 Gyr.  Whatever
events produced the `high-alpha' and `low-alpha' halo stars occurred during the first few Gyr
of the Galaxy's formation.  The $\Lambda$CDM simulations of Abadi et al.\ (\cite{abadi03a},
\cite{abadi03b}) for disk galaxies similar to the \object{Milky Way} also show that only a
very few ($\la 5$) major merger or accretion events occurred, and quite early, producing stellar
ages today greater than nine Gyr (see their Figs.~5 and 7, respectively).  Then, were these
halo progenitors large enough, $\sim 5\times 10^{10} M_{\odot}$, to form the `high-alpha'
halo stars, as in Fig.~3 of Robertson et al.\ (\cite{Robertson05}), or did the gas merely
combine to form a primeval bulge or disk massive enough to form the `high-alpha' halo stars
with subsequent displacement into the inner halo by the same later mergers which populated the
outer halo $\grave{a}\ la$ Zolotov et al.\ (\cite{zolotov09}, \cite{zolotov10}) or Purcell et al.\ 
(\cite{purcell10})?  The `low-alpha' halo stars then represent the remains of more modest
accretion events which took place a couple Gyr after the stars of the `high-alpha' halo, i.e.\ 
the `low-alpha' halo stars constitute tidal debris (Navarro et al.\ \cite{navarro11}).


\begin{figure}
\resizebox{\hsize}{!}{\includegraphics{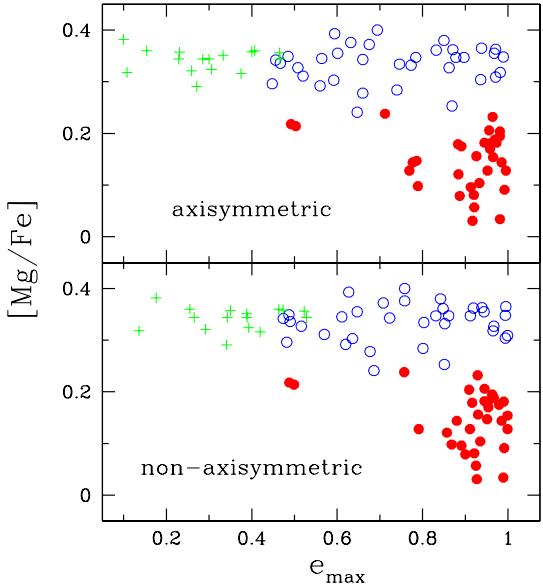}}
\caption{The `high-alpha' halo, `low-alpha' halo, and thick-disk
stars are plotted in the \mgfe\ vs ${\rm e}_{\rm max}$ diagram for
orbital integration times of 5 Gyr and for the axisymmetric Galactic
potential in the upper panel, and non-axisymmetric in the lower.
The symbols as in Fig.~\ref{fig:logg-logteff1}.}
\label{fig:mgfe-e1max.2}
\end{figure}

\begin{figure}
\resizebox{\hsize}{!}{\includegraphics{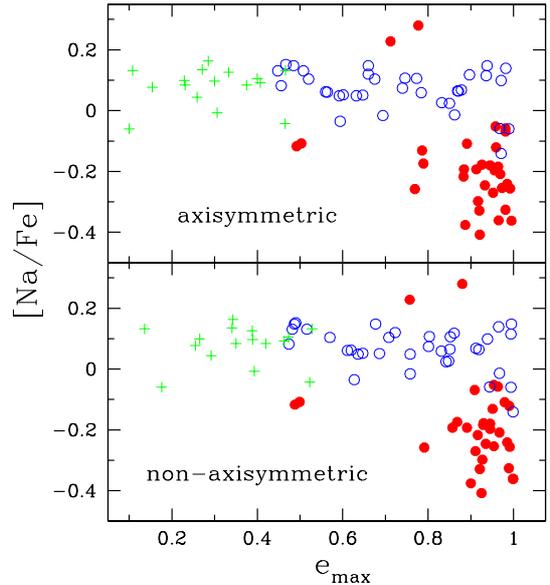}}
\caption{The `high-alpha' halo, `low-alpha' halo, and thick-disk
stars are plotted in the \nafe\ vs ${\rm e}_{\rm max}$ diagram for
orbital integration times of 5 Gyr and for the axisymmetric Galactic
potential in the upper panel, and non-axisymmetric in the lower.
The symbols as in Fig.~\ref{fig:logg-logteff1}.}
\label{fig:nafe-e1max.2}
\end{figure}


The ELS-plus-SZ or dual-accretion scenarios (Robertson et al.\ \cite{Robertson05};
Font et al.\ \cite{font06a}) have difficulty explaining, or producing, high-alpha halo stars;
mainly they produce one or more sequences of low-alpha halo stars.  Perhaps the
dual-accretion models might be salvaged using a very few, very massive accretion components
which are merged and destroyed quickly to form the primeval Galaxy and its first halo
component.

\begin{figure}
\resizebox{\hsize}{!}{\includegraphics{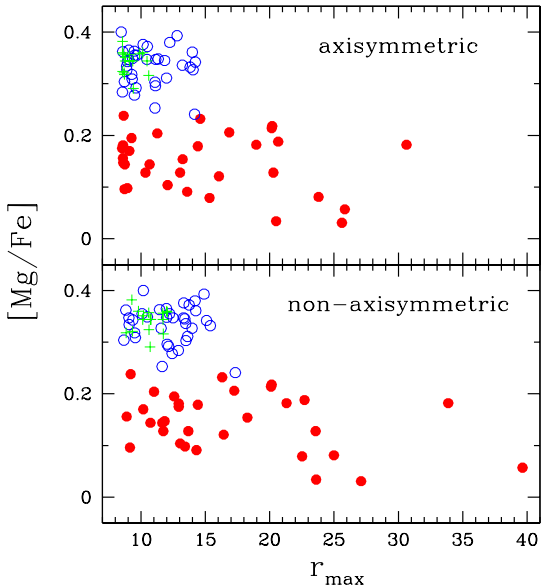}}
\caption{The `high-alpha' halo, `low-alpha' halo, and thick-disk
stars are plotted in the \mgfe\ vs $r_{\rm max}$ diagram for
orbital integration times of 5 Gyr and for the axisymmetric Galactic
potential in the upper panel, and non-axisymmetric in the lower.
The symbols as in Fig.~\ref{fig:logg-logteff1}.}
\label{fig:mgfe-rmax.2}
\end{figure}

\begin{figure}
\resizebox{\hsize}{!}{\includegraphics{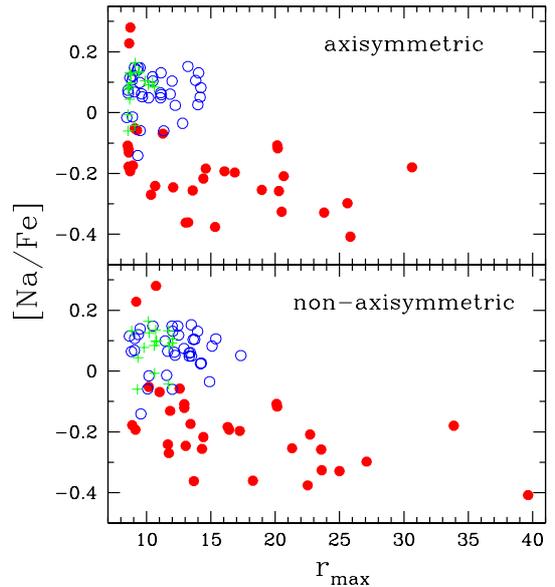}}
\caption{The `high-alpha' halo, `low-alpha' halo, and thick-disk
stars are plotted in the \nafe\ vs $r_{\rm max}$ diagram for
orbital integration times of 5 Gyr and for the axisymmetric Galactic
potential in the upper panel, and non-axisymmetric in the lower.
The symbols as in Fig.~\ref{fig:logg-logteff1}.}
\label{fig:nafe-rmax.2}
\end{figure}

The orbital parameters from the integrations discussed above for the symmetric and non-symmetric
Galactic potentials give some additional very important clues concerning the origins of these
two halo components detected here in the Solar vicinity.  For example, in 
Figs.~\ref{fig:mgfe-rmax.2} and \ref{fig:nafe-rmax.2},
the abundances \mgfe\ and \nafe , respectively, are plotted as a function of
the maximum distances from the Galactic center reached by these stars in the last 5 Gyr,
with the same symbols as in  Fig.~\ref{fig:logg-logteff1}.  And, in 
Figs.~\ref{fig:mgfe-e1max.2} and \ref{fig:nafe-e1max.2},
the same abundances with the same symbols, respectively, are plotted as a function of the
maximum orbital eccentricities reached by these stars in 5 Gyr.  In all of the plots versus
${\rm r}_{\rm max}$ the `low-alpha' halo stars reach much greater distances from the Galactic
center than the `high-alpha' halo stars, and in all four of these plots there seems to be a
correlation of decreasing \mgfe\ or \nafe\ values with increasing 
${\rm r}_{\rm max}$ for the `low-alpha' halo stars.  The `high-alpha' halo stars reach a
limiting ${\rm r}_{\rm max}$ value of about 16 kpc, while the `low-alpha' halo stars go out
as far as 30--40 kpc in some cases.  In the plots versus the orbital eccentricities,
`high-alpha' halo stars have on the average more circular orbits with a more or less uniform
distribution over $0.4 \la {\rm e}_{\rm max} \la 1.0$, while the `low-alpha' ones are largely
clumped at ${\rm e}_{\rm max} \ga 0.85$ with only a few exceptions.  One might conclude that
these results support the ideas of Zolotov et al.\ (\cite{zolotov09}, \cite{zolotov10}), since 
the `high-alpha' halo stars populate exclusively the inner halo, (${\rm r}_{\rm max} \la 16$),
and exhibit a wider, uniform range in orbital eccentricities, which might be expected for stars
kicked out of a primeval bulge or primeval disk by merger or accretion events.

However, most ($\ga 80\%$) of the $in \: situ$ halo stars of Zolotov et al.\ (\cite{zolotov09},
\cite{zolotov10}) reside within the inner $\sim10$ kpc of their simulated halos (see Fig.~2 of
Zolotov et al.\ \cite{zolotov09}, or Sect.~2.1 of Zolotov et al.\ \cite{zolotov10}), contrasting
with our limit here of about 16 kpc for the `high-alpha' halo stars.  In addition, it is not
clear whether the \alphafe\ abundances of stars now residing in the Galactic Bulge are higher
than those of the thick disk (Zoccali et al.\ \cite{zoccali06}; Lecureur et al.\ \cite{lecureur07}),
or the same as the thick disk (Gonzalez et al.\ \cite{gonzalez11}), and in our abundance diagrams
the thick-disk stars follow, or extend uniformly, the abundances of the `high-alpha' halo stars.
These facts would suggest that the displacement of $in \: situ$ stars into the inner halo, as
proposed by Zolotov et al.\ (\cite{zolotov09}, \cite{zolotov10}), did not occur exclusively from the
primeval bulge but also from a primeval disk, or thick disk, which extended out to nearly 16 kpc
from the center of the Galaxy (Purcell et al.\ \cite{purcell10}).


The \ofe\ versus \feh\ diagrams from Zolotov et al.\ (\cite{zolotov10}; Fig.~3) fit fairly well
the \alphafe\ diagrams of our Paper I (especially their models MW1 and h277), and they conclude
that a galaxy which has had a long quiescent period, $\approx 9$ Gyr, without any major mergers,
should have a fairly high proportion of $in \: situ$ halo stars ($\approx 20 \%$--$50 \%$) as
compared to the percentage of halo stars from mergers or accretions.  A galaxy with many more recent
mergers might have an $in \: situ$ halo contribution as low as 5\%.  Our sample includes 35 (52\%)
`high-alpha' halo stars and 32 (48\%) of the `low-alpha' halo stars, all from the solar neighborhood,
and so these percentages for our `high-' and `low-alpha' halo stars agree with their large ages,
10--13 Gyr, qualitatively at least, according to these models.  As mentioned above, our age
determinations confirm that the Galaxy seems to have had a rather long quiescent period, without any
major mergers, after the first events which formed the primeval Galaxy.

Dual-accretion $\lambda$CDM models for the Galaxy (Font et al.\ \cite{font06a}; Robertson et al.\
\cite{Robertson05}) in general do not fit as well the details of the \alphafe , or \mgfe , versus
\feh\ diagrams of Paper I.  See for example, Fig.~9 of Font et al.\ (\cite{font06a}), where two
low-alpha sequences are produced but no high-alpha one, or Fig.~3 of Robertson et al.\
(\cite{Robertson05}), with a similar problem.  In this latter paper, of the trichotomy of models
(``halo progenitor'', dSph, and dIrr) the first seems to come closest to producing `high-alpha'
halo stars.  So, this model might be modified by increasing the mass of the progenitors and merging
very quickly at the beginning of the Galaxy, combined with a dIrr model, to produce the dichotomy in
the \alphafe\ versus \feh\ diagrams of Paper I, both the `high-' and `low-alpha' halo stars.  (See
also De Lucia \& Helmi \cite{delucia08} for their discussion of the ``duality'' in their modeled
stellar halo.)  However, the range in age for the `high-alpha' halo stars, as seen in Table 2, is
probably larger than that needed for such a scenario, especially if those in the highest metallicity
range, $-0.575 \le [\rm {Fe/H}] \le -0.40$, are considered.

Our relative-age sequence, with the `low-alpha' halo stars being 2--3 Gyr younger than the `high-alpha'
halo stars, and the thick-disk ages lying in between, is probably congruous with any of the model-types
mentioned above, ELS-plus-SZ, $in \: situ$ plus accretion, or dual-accretion, for the formation of the
Galactic halo.  In the models of Zolotov et al.\ (\cite{zolotov09}, \cite{zolotov10}) infalling gas
forms the primeval components in which the $in \: situ$ halo stars are formed, and then 2-3 Gyr later
merger or accretion events kick these $in \: situ$ `high-alpha' stars into the inner halo and at the
same time populate the `low-alpha' component of the halo.  For the dual-accretion $\lambda$CDM models
for the Galaxy, a very few quick and massive accretion events might produce the `high-alpha' component,
and then 2-3 Gyr later a few major accretion events the `low-alpha'' halo stars.  The relative
kinematics of these competing models for the Galactic halo are not well documented in these papers.

Recently, Font et al.\ (\cite{font11}) have carried out new simulations of the formation of stellar
halos around disk galaxies.  The simulations, which include baryons in a self-consistent way, show that
$in \: situ$ star formation dominates in the inner halo ($r < 30$\,kpc), whereas most stars in the
outer halo have been accreted from satellite galaxies.  An apparent gradient in [Fe/H] is
predicted due to changing proportions of the $in \: situ$ and accreted components; the
metallicity drops by 0.6 - 0.9 dex from the inner to the outer halo in agreement with the analyses of
SDSS data by Carollo et al.\ (\cite{carollo10}) and de Jong et al.\ (\cite{dejong10}).  In contrast to
Zolotov et al.\ (\cite{zolotov09}), who found that the $in \: situ$ stars formed very early out of
accreted cold gas in the inner 1 kpc region and were later displaced to the halo by mergers, Font et al.\
(\cite{font11}) find that $in \: situ$ halo stars formed from cooling of hot gas in a more extended
inner region and were, on average, not displaced significantly from their formation site.  Furthermore,
they predict that the stars formed $in \: situ$, with ages $\la 8$ Gyr, are 3-4 Gyr younger than the
accreted stars.  From this it is clear that we cannot readily identify our `low-alpha' and `high-alpha'
halo stars with the populations of accreted and $in \: situ$ stars in the Font et al.\ (\cite{font11})
simulations.  The `low-alpha' stars are about 2 Gyr younger than the `high-alpha' stars (presumably
formed $in \, situ$), and they belong to the metal-rich end of the metallicity distribution, whereas
the accreted stars of Font et al.\ (\cite{font11}) belong to the low-metallicity end.  And, our
`high-alpha' halo stars have ages in the range 11--14 Gyrs, and orbital eccentricities, 0.4--1.0.  A
more detailed study is needed to see if the Font et al.\ (\cite{font11}) simulations can produce two
sequences of \alphafe\ vs. \feh\ for metal-rich halo stars that match the age and orbital characteristics
of our two halo components.

The probable correlations of the abundances \mgfe\ and \nafe\ with
${\rm r}_{\rm max}$ for the `low-alpha' halo stars are very interesting and arresting, and
may provide an important clue as to their origin.  (These correlations are significant at
the 3--4$\sigma$ level for \mgfe\ and 9--10$\sigma$ for \nafe, excluding the two outliers, where
10\% errors in ${\rm r}_{\rm max}$ and the abundance errors in \mgfe\ and \nafe\ from Paper I
have been included.)  For example, these stars have probably come from
progenitors falling into the Galaxy on nearly radial orbits, as shown by the large orbital
eccentricities  (${\rm e}_{\rm max} \ga 0.85$) and low angular momenta (as seen in the Toomre
diagram, Fig.~\ref{fig:toomre}, especially if ${\rm V}_{\rm LSR} = 254$ \kms, as in the orbital
integrations).  The progenitors will oscillate almost radially in the Galactic potential with
their maximum distances from the Galactic center slowly decreasing due to dynamic friction.  At
the same time these progenitors will undergo internal chemical evolution and will suffer tidal
stripping each time they pass through the Galactic bulge or disk.  So with time the
${\rm r}_{\rm max}$'s of the progenitors will decrease, as will their total mass and their
ability to retain the chemical output from the SNeIa and SNeII.  At first, with higher mass,
they will retain at least some of the SNeIa remnants and will contribute `low-alpha' stars to
the field with each tidal stripping.  Later when their total mass has been reduced considerably
(more than a factor of ten as has been suggested for some progenitors (Schaerer \& Charbonnel
\cite{schaerer11}), the outer zones where such SNeIa remnants have been retained will be gone, and
these progenitors will contribute mainly older `high-alpha' stars to the field.  So, according to
this scenario, the dynamic time lines for the `low-alpha' halo stars in Figs.~\ref{fig:mgfe-rmax.2}
and \ref{fig:nafe-rmax.2} actually run opposite that of the chemical evolution, from right to left,
rather than left to right, from large ${\rm r}_{\rm max}$ and large mass for the progenitor, on the
right, to smaller values of each on the left.  

A case in point is the globular cluster \object{$\omega$ Cen}, which is thought to have been a dwarf galaxy
accreted by the Galaxy several Gyr ago.  Results by Dinescu (\cite{dinescu02}) have shown metal-poor stars
in the solar neighborhood with a metallicity range including that of \object{$\omega$ Cen}, with a
retrograde signature similar to \object{$\omega$ Cen}'s orbit, and with many orbital eccentricities larger
than 0.8.$\;$ \object{$\omega$ Cen} appears to have undergone considerable internal chemical evolution
with at least four peaks in the [Fe/H] distribution (Johnson \& Pilachowski \cite{Johnson10}),
an internal age spread of 3--8 Gyr (Hilker \& Richtler \cite{hilker00}, \cite{hilker02}; Hughes \&
Wallerstein \cite{hughes00}; Hughes, Wallerstein, \& van Leeuwen \cite{hughes02}; Rey et al.\ \cite{rey02};
Smith et al.\ \cite{smith00}; Smith \cite{smith02}, and references therein), and, despite this significant
age spread, almost no evidence of contributions from Type Ia SNe (Johnson \& Pilachowski \cite{Johnson10};
Smith et al.\ \cite{smith00}; and Norris \& Da Costa \cite{norris95}); only a very few
low-alpha--type stars have perhaps been found in \object{$\omega$ Cen} (Pancino et al.\ \cite{pancino02}).
It has been estimated that \object{$\omega$ Cen} has lost more than three-fourths of its original mass
(Bekki \& Freeman \cite{bekki03}), perhaps more than 99\% (Dinescu et al.\ \cite{dinescu99b}, Freeman \&
Bland-Hawthorn \cite{freeman02}), and so no longer retains the SNeIa ejecta nor the zone in which low-alpha
stars were produced in the past, and therefore \object{$\omega$ Cen} no longer produces nor contributes
low-alpha halo stars to the field.  But with a much higher mass in the past, it may have done so with these
SNeIa ejecta accumulating in an outer shell of the progenitor.  In Paper II a comparison between the
`low-alpha' halo stars and \object{$\omega$ Cen} was made due to a similarity in the (U,V,W) Galactic
velocities, and conclusions drawn similar to those above due to the lack now of any significant number of
stars showing enrichment by SNeIa ejecta.  The above scenario explains how \object{$\omega$ Cen} may have
contributed significantly to the `low-alpha' halo stars, and yet nowadays have so very few of these.

Our Galactic mass model, with the bar only, and using the measured absolute proper motions for $\omega$
Cen from Dinescu et al.\ (\cite{dinescu99a}, \cite{dinescu99b}), has been used to integrate its orbit for
5 Gyr, producing an apogalactic distance of about 7.5 kpc, a perigalactic distance of about 0.5 kpc,
${\rm z}_{\rm max} \approx 3.7$ kpc, and ${\rm e}_{\rm max}\approx 0.87$.  This eccentricity agrees with
those of the `low-alpha' halo stars in Fig.~\ref{fig:e1max-rmax6}, but these ${\rm r}_{\rm max}$ and
${\rm z}_{\rm max}$ values fall far short as seen in Fig.~\ref{fig:zmax-rmax6} with values of
${\rm r}_{\rm max}$ out to 30-40 kpc and ${\rm z}_{\rm max}$ values generally above 5 kpc for the `low-alpha'
halo stars.   Dinescu et al.\ (\cite{dinescu99b}) also used these absolute proper motions to integrate
Galactic orbits for \object{$\omega$ Cen} with two different Galactic mass models giving an apogalactic
distance of about 6.2 kpc and a perigalactic distance of about 1.2 kpc, with  ${\rm z}_{\rm max} \approx 1.0$
kpc and ${\rm e} \approx 0.67$.  These values agree even less with the typical orbital parameters of the
`low-alpha' halo stars.  

Simulations by Zhao (\cite{zhao02}) would suggest that \object{$\omega$ Cen} would have had to been
launched near the edge of the Galactic disk with R $\approx$ 15 kpc, $|{\rm Z}| \approx$ 1 kpc, and a
retrograde velocity to obtain its observed mass and apo- and peri-galactic distances in about a Hubble time;
these simulations take into account dynamical friction and tidal stripping, unlike the orbital
integrations mentioned above.  Again these initial conditions and orbital characteristics are not
in good agreement with our orbital parameters for many of the `low-alpha' halo stars.  The
simulations of Bekki \& Freeman (\cite{bekki03}) for the orbit and tidal stripping of \object{$\omega$ Cen}
might yield accreted stars out to about R = 26 kpc in 2.6 Gyr, and produce approximately the current
apogalactic distance of \object{$\omega$ Cen}, 8 kpc, which compares well with the 7.5 kpc value obtained
with our Galactic mass model.  That simulation which approximately fits the current orbital
parameters of \object{$\omega$ Cen}, and at the same time spans well our `low-alpha' halo stars, has been
obtained by Tsuchiya et al.\ (\cite{tsuchiya03}, \cite{tsuchiya04}) by providing \object{$\omega$ Cen} with
M$_{tot}$ = $8\times 10^{9}$ M$_{\odot}$ at the beginning of the integrations, a Hernquist density
profile (their model H4), a launching distance of R = 58 kpc., i.e.\ 50 kpc from the rotation
axis and 30 kpc above the Galactic plane, with a retrograde velocity of $-20$ \kmprs in the direction
of Galactic rotation, and zero velocities in the other two directions; they obtain current apogalactic
and perigalactic distances of about 6 and 1 kpc, and a remaining mass for \object{$\omega$ Cen} of
approximately $10^7$ M$_{\odot}$ after about 3 Gyr.  Their method improves on simplifications made by Zhao
(\cite{zhao02}); for example, they have also included dynamical friction from the Galactic bulge and disk.
This model could produce stripped stars like the `low-alpha' halo stars, with ${\rm r}_{\rm max}$'s to 40 kpc,
${\rm z}_{\rm max}$'s greater than 5 kpc, and very eccentric orbits.

So, if \object{$\omega$ Cen} has made a significant contribution to our `low-alpha' halo stars, this would
imply that it  has probably suffered very significant dynamical friction and tidal stripping.
Some of the above models estimate that it has lost more than 99\% of its original mass, for example,
from $8\times 10^9$ M$_{\odot}$ to about $1\times 10^7$ M$_{\odot}$ in the models of Tsuchiya et
al.\ (\cite{tsuchiya03}, \cite{tsuchiya04}).  This would help explain why and how \object{$\omega$ Cen}
now has almost no stars with evidence of contributions from Type Ia SNe (Johnson \& Pilachowski
\cite{Johnson10}; Smith et al.\ \cite{smith00}; and Norris \& Da Costa \cite{norris95}), but may have
contributed in the past a significant portion of the `low-alpha' halo stars now seen in the solar
neighborhood.  The type Ia SNe remnants were deposited mostly in the outer regions of its progenitor,
and these have by now been stripped away.  In analogy, significant differences between the abundances of
the Sagittarius dwarf galaxy and its stream have been detected by Chou et al.~(\cite{chou07}, \cite{chou10}).

Due to the large ages with a moderately small dispersion of the `low-alpha' halo stars, it is
improbable that the Sagittarius dwarf galaxy has contributed to their presence in the solar
neighborhood; this dwarf galaxy is most likely on one of its first passes near the \object{Milky Way}
(Ibata et al.\ \cite{IG95}) and contains a mixture of stars with ages from about 6 Gyr to more than 9 Gyr
(Bellazzini et al.\ \cite{bellazzini06}).  In agreement, both local radial velocity surveys and
calculations of this dwarf galaxy's orbit have concluded that it has not passed through the solar
neighborhood (for example, Seabroke et al.\ \cite{seabroke08}; Newberg et al.\ \cite{newberg07}).

Very recent results by Tan \& Zhao (\cite{tan11}) supplement and support the results of our Papers I,
II, and III.  They have obtained beryllium abundances, on a uniform scale, for 43 of our halo
and thick-disk stars, including 14 thick-disk stars, 13 `low-alpha', and 16 `high-alpha' halo
stars.  They find very clear evidence for separate sequences between the `low-alpha' and
`high-alpha' (halo plus thick-disk) stars in the A(Be) versus \feh, and A(Be) versus
[$\alpha$/H] diagrams, with the `high-alpha' stars having larger A(Be) values for a given
value of \feh, or [$\alpha$/H].  Since Be is formed by cosmic-ray reactions, these results
would suggest that these two stellar components have formed in regions with different strengths
of the cosmic-ray field, such as a low-mass dwarf galaxy versus a high-mass primeval bulge or
disk.

From this discussion of our results, we draw the following conclusions about stars
belonging to the metal-rich end ($\feh > -1.6$) of the halo metallicity distribution.

\begin{enumerate}

   \item The ages of both the `high-' and `low-alpha' halo stars are large, 10-13 Gyr, supporting
         the idea, presented in many works in the literature, that the Galaxy has had a rather
         quiescent history for the last 9 Gyr, or more, without any major mergers.  Also, our ages
         from the Y$^2$ isochrones, interpolated to the exact \feh\ and \alphafe\ values of the
         spectroscopic analyses, indicate that the `high-alpha' halo stars are 2-3 Gyr older than
         the `low-alpha', with the possibility that the thick-disk stars have ages intermediate
         between these two halo components.

   \item The $in \: situ$-accretion models of Zolotov et al.\ (\cite{zolotov09}, \cite{zolotov10}) can
         explain more naturally both the `high-' and `low-alpha' halo stars.  Infalling gas forms a
         primeval bulge or disk which in turn produce $in \: situ$ stars that are then scattered into the
         inner halo by the same accretion events which bring into the Galaxy the `low-alpha' halo stars.
         Both $in \: situ$ and accreted stellar products end up in the Galactic halo, with greatly differing
         but overlapping distributions, as seen in the orbital parameters of our two halo components,
         and these are found in our halo sample with nearly equal numbers in agreement with the recent
         quiescent history of the \object{Milky Way}.

   \item The distribution of orbital eccentricities is quite different between the `high-' and `low-alpha'
         halo stars; see Figs.~\ref{fig:e1max-rmax6}--\ref{fig:nafe-e1max.2}.  This supports the idea of
         different formation scenarios for these two components, such as scattered $in \: situ$ halo
         stars versus accreted halo stars, respectively, with the thick-disk being an extension of the
         `high-alpha' halo, or vice versa.

   \item The distributions of ${\rm r}_{\rm max}$ and ${\rm z}_{\rm max}$ are also significantly different
         between the `high-' and `low-alpha' halo stars; see Fig.~\ref{fig:zmax-rmax6}.  The `low-alpha'
         ones reach out to ${\rm r}_{\rm max} \approx$ 30--40 kpc, while the `high-alpha' only to
         ${\rm r}_{\rm max} \approx$ 16 kpc, and for ${\rm z}_{\rm max}$, the `low-alpha' halo stars
         reach much higher levels, $\approx$ 18 kpc, than the `high-alpha' ones, $\approx$ 6-8 kpc;
         these latter values are in line with the heights produced by scattering from a primeval disk in
         the simulations of Purcell et al.\ (\cite{purcell10}).  However, the $in \: situ$ halo stars of
         the models are in general scattered from the bulge into the inner halo with ${\rm r}_{\rm max}$'s
         less than 10 kpc; more than 80\% have ${\rm r}_{\rm max} < 10$ kpc, suggesting that more energetic
         scattering from the primeval bulge is required in the models of Zolotov et al.\ (\cite{zolotov09},
         \cite{zolotov10}).

   \item Significant correlations between the abundances \mgfe\ or \nafe\ and the orbital
         parameter ${\rm r}_{\rm max}$ are seen in Figs.~\ref{fig:mgfe-rmax.2} and \ref{fig:nafe-rmax.2}
         for the `low-alpha' halo stars.  We suggest that these correlations have to do with the
         combined effects of tidal stripping and dynamical friction on the progenitors of the `low-alpha'
         halo stars, combined with internal chemical evolution and radial and temporal variations in the
         retention of SNeIa remnants.  Such a scenario fits well the suggestion that \object{$\omega$ Cen}
         has contributed in a significant way to the `low-alpha' halo stars.

   \item The models of Tsuchiya et al.\ (\cite{tsuchiya03}, \cite{tsuchiya04}) give a reasonable fit to
         the current mass and orbital characteristics of \object{$\omega$ Cen}, while at the same time
         providing a feasible explanation for the large values of ${\rm r}_{\rm max}$, ${\rm z}_{\rm max}$,
         and ${\rm e}_{\rm max}$ of the `low-alpha' halo stars as seen in Figs.~\ref{fig:zmax-rmax6} and
         \ref{fig:e1max-rmax6}.  Their orbital integrations were begun with the progenitor 58 kpc from
         the Galactic center, and with a retrograde rotational velocity of $-20$ \kmprs.  Such initial
         conditions are probably representative of other dwarf galaxies which may have contributed to
         the `low-$\alpha$' halo stars.

\end{enumerate}

As epilogue, it seems that evidence in the \object{Milky Way} is ubiquitous, and probably in many other galaxies as well,
for some sort of dominating, quick event at the very beginning of galaxy formation, such as a rapid monolithic
collapse of a protogalactic cloud \'a la ELS, which would have produced the high-$\alpha$ halo stars.  This
evidence is seen in the main sequence turnoffs as studied in the SDSS data by Jofr\'e \& Weiss (\cite{Jofre11})
and in the $(b$--$y)_{\rm 0,TO}$ versus \feh\ diagram for high-velocity and metal-poor halo stars of Schuster et al. 
(\cite{schuster06}).  This evidence is also clear in the old-group globular clusters of Mar\'in-Franch et al.
(\cite{marinfranch09}) with a mean age of about 12.8 Gyr and an intrinsic dispersion of only about 0.4 Gyr, and in
the older population of numerous early-type galaxies encountered by Rakos et al. (\cite{rakos08}) with ages of 11--12
Gyr.  Such a quick, significant, and early event would also be needed to produce the high-$\alpha$ halo stars of
this paper with metallicities as high as [Fe/H] $\approx -0.4$ before the ignition of SNeIa, and ages as large as
11--12 Gyr.  This goes whether or not they were collapsed, merged, or accreted directly into the Galactic halo,
or scattered there.   However, such a quick and early event is not compatible with hierarchical galaxy-formation
schemes which require the gradual merging of smaller units to form large galaxies.

Also, the low-$\alpha$ halo stars pose a quandary, together with the results of Mar\'in-Franch et al.
(\cite{marinfranch09}) for their young-group, probably-accreted, globular clusters.  In Paper II of this series
it has been shown that the dispersions of the low-$\alpha$ sequences are larger than the expected observational
spectroscopic errors with values of 0.04--0.06 dex for the $\alpha$ elements, and dispersions as large as
0.11--0.15 dex for [Na/Fe] and [Cu/Fe].  It was suggested that this is due to the accretion of an ensemble of
dwarf galaxies with slightly different star-formation histories.  But the question arises: why is this scatter
not much larger?  Figure 4 of Zolotov et al. (\cite{zolotov10}) would suggest a much wider low-$\alpha$
sequence, perhaps 0.3 dex or more in \alphafe\ at a given \feh , for any significant and reasonable range for
the masses of these accreted dwarf galaxies.  And, the \alphafe\ versus \feh\ diagrams of Tolstoy et al.\ (2009;
Fig.~11) for high-resolution spectroscopy of giant stars in four dwarf spheroidal galaxies would suggest a range as
large as 0.3 dex (e.g. between Sculptor, Fornax, and Sagittarius).   In this same context, Mar\'in-Franch et al.\ 
(\cite{marinfranch09}) question why all of their young-group accreted globular cluster follow the same
age-metallicity sequence (i.e.\ similar to an [$\alpha/$Fe] versus [Fe/H] sequence) if they have been accreted
from different dwarf galaxies, such as Sagittarius, Monoceros, and Canis Major?  Does this indicate that all
previous dwarf galaxies shared very similar star-formation and chemical-enrichment histories, which as Mar\'in-Franch
et al.(\cite{marinfranch09}) point out seems very ``unlikely''!  Why should all of these dwarf galaxies (or their
globular clusters) have had similar initial masses?

\begin{acknowledgements}
This publication made use of the SIMBAD database operated
at CDS, Strasbourg, France, and of data products from the Two Micron All
Sky Survey, which is a joint project of the University of Massachusetts and
the Infrared Processing and Analysis Center/California Institute of
Technology, funded by NASA and the National Science Foundation.
This research has made use of NASA's Astrophysics Data System.
The staff at the Nordic Optical Telescope is thanked for competent
and friendly assistance in obtaining spectra for this project, and
the referee, Timothy Beers, for constructive and useful suggestions.
\end{acknowledgements}

\Online
\end{document}